\documentclass[a4paper,twocolumn,preprintnumbers,citeautoscript,aps,prd,longbibliography,superscriptaddress,floatfix,nofootinbib]{revtex4-2}
\usepackage{amsmath,amsfonts,amssymb,graphics}
\usepackage[normalem]{ulem}
\usepackage{url}
\usepackage{graphicx}

\usepackage{mathrsfs}
\usepackage{bbm}
\usepackage{pifont}

\usepackage{siunitx}

\usepackage{mathtools}
\usepackage{xspace}
\usepackage{booktabs}
\usepackage{csquotes}
\usepackage[dvipsnames,svgnames]{xcolor}
\usepackage{tikz}
\usetikzlibrary{positioning,shapes,arrows}
\usetikzlibrary{decorations.pathmorphing}
\usetikzlibrary{decorations.markings}
\usepackage{standalone}

\usepackage{url}
\usepackage[bookmarks=false, breaklinks]{hyperref} 
\usepackage{braket}

\definecolor{nicered}{rgb}{0.5,.0,.0}
\definecolor{darkblue}{rgb}{0,.1,.9}
\definecolor{lightblue}{rgb}{0,.1,.6}
\definecolor{applegreen}{rgb}{0.55, 0.71, 0.0}
\definecolor{darkgreen}{rgb}{0.0, 0.2, 0.13}
 
\hypersetup{colorlinks, citecolor=darkgreen ,linkcolor=nicered, urlcolor= lightblue}

\begin{document}
\title{The Type II Dirac Seesaw Portal to the mirror sector:\\ Connecting neutrino masses and a solution to the strong CP problem}
\author{Maximilian Berbig}
\email[]{berbig@physik.uni-bonn.de}
\affiliation{Bethe Center for Theoretical Physics und Physikalisches 
Institut der Universit\"at Bonn, \\ Nussallee 12, 53115 Bonn, Germany}
\date{\today}

\begin{abstract}
    \noindent We present a   version of the Type II Seesaw mechanism for parametrically small Dirac neutrino masses. Our model starts from an $\text{SU}(2)_\text{L} \otimes \text{SU}(2)'\otimes \text{U}(1)_\text{X}$ gauge extension of the Standard Model  involving a sector of mirror fermions. A bidoublet scalar with a very small vacuum expectation value connects the SM leptons with their mirror counterparts  and we can identify the mirror neutrino with the right-handed neutrino. Similar to the conventional  Type II Seesaw, our particle spectrum features singly- and doubly-charged scalars. The strong CP problem is solved by a discrete exchange symmetry between the two sectors that forces the contributions of quarks and mirror quarks to the strong CP phase to cancel each other.  We discuss the low-energy phenomenology, comment on the cosmological implications of this scenario and indicate how to realize successful Dirac leptogenesis.
\end{abstract}

\maketitle

\section{Introduction}
\begin{figure}[t]
 \centering
  \includegraphics{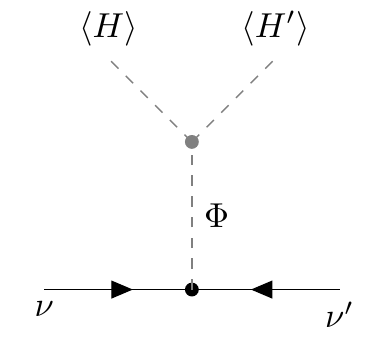}
  \caption{Diagrammatic representation of the Type II Dirac Seesaw mechanism. $\nu\;(\nu')$ is embedded in the doublet $l\;(l')$. The mirror neutrino $\nu'$ plays the role of the right-chiral neutrino and the heavy scalar $\Phi$ is integrated out. }
  \label{fig:numass}
\end{figure}

\noindent The Type II Seesaw mechanism \cite{Lazarides:1980nt,Schechter:1980gr,Mohapatra:1980yp,PhysRevD.22.2860,Wetterich:1981bx} offers an approach to parametrically small neutrino masses, that is somewhat orthogonal to Type I Seesaw schemes involving fermionic messengers \cite{Yanagida:1979as,10.1143/PTP.64.1103,Minkowski:1977sc,Gell-Mann:1979vob, PhysRevLett.44.912}. For Majorana neutrino masses, one needs to incorporate a weak isotriplet field with hypercharge $-1$. One finds that the vacuum expectation value (vev) $v_\Delta$ of the super-heavy scalar triplet $\Delta$ with mass $\mu_\Delta^2$ is induced by the vev $v$ of the Standard Model (SM)-like Higgs $v_\Delta\simeq \kappa v^2 /\mu_\Delta^2$, where $\kappa$ is the dimensionful coupling for the term $\kappa H\Delta H$.
This occurs because after $H$ gets a vev, the aforementioned term biases the triplet potential in one direction  such that a non-trivial minimum can appear despite the fact that $\mu_\Delta^2 >0$. Since such a triplet breaks the custodial symmetry of the SM scalar potential, its vev is tightly constrained to lie below the GeV scale \cite{Kanemura:2012rs} by the observed ratio of the $W^\pm$ and $Z$ boson masses encoded in the $\rho$-parameter \cite{Veltman:1976rt}. A typical feature of such scalar extensions for either  tree- or loop-level Majorana neutrino masses \cite{Zee:1985id,Babu:1988ki,Gustafsson:2012vj} is the presence of double charged scalars whose production at colliders provides a smoking gun signature for these scenarios. As so far there is no compelling experimental or theoretical indication to consider only Majorana neutrinos, the field of Dirac model building has received renewed attention in the last years. It is possible to construct  Dirac equivalents of all conventional Seesaw mechanisms \cite{Ma:2016mwh,CentellesChulia:2018gwr}. The usual approach for light Dirac neutrino masses is to start with a symmetry forbidding the tree-level mass term from the Standard Model (SM) Higgs doublet. Small neutrino masses at tree-level can then be realized by inducing a small vev for a  new Higgs doublet \cite{Ma:2000cc,Bonilla:2016zef,Bonilla:2017ekt,Gu:2019ird}. In gauge extensions of the SM like the left-right symmetric model (LRSM) $\text{SU}(2)_\text{L} \otimes \text{SU}(2)_\text{R}\otimes \text{U}(1)_\text{B-L}$  \cite{Pati:1974yy,Mohapatra:1974gc,Senjanovic:1975rk,Mohapatra:1979ia,Deshpande:1990ip} or \textit{mirror-sector} constructions \cite{Foot:1991py,Foot:1995pa}, the role of the doublet can be played by a bidoublet field \cite{Gu:2012fg,Gu:2013nya,Gu:2013vpx}, whereas in $\text{SU}(3)_\text{C} \otimes \text{SU}(3)_\text{L}\otimes \text{U}(1)_\text{X}$ models \cite{Singer:1980sw} Higgs triplets of $\text{SU}(3)_\text{L}$ are needed \cite{Valle:2016kyz,Reig:2016ewy}.
In contrast to the Majorana case with its doubly-charged scalars, these scenarios feature only singly-charged ones, which are quite generic from a beyond the Standard Model (BSM) perspective. In this work we set out to close this gap by constructing a Dirac neutrino mass model with the same singly- and doubly-charged scalar spectrum as the original Type II Seesaw. The new scalar will also play an important role for leptogenesis. Our starting point is the original mirror fermion scenario of reference \cite{PhysRevLett.67.2765} that was designed to solve the strong CP problem with the help of a discrete exchange symmetry. In recent years it was shown \cite{Hall:2018let,Dunsky:2019api} that one does not need to copy the entire $\text{SU}(2)\otimes \text{U}(1)$ structure of the SM, but that the gauge group of the form $\text{SU}(2)_\text{L} \otimes \text{SU}(2)'\otimes \text{U}(1)_\text{X}$ similar to the LRSM is already enough.
While the  proposal \cite{Hall:2018let,Dunsky:2019api} used additional singlet fermions to form a Type I Seesaw, our approach is to include the following bidoublet scalar
\begin{align}\label{eq:bivev}
    \Phi = \begin{pmatrix}\varphi_1^0 & \varphi_2^- \\ \varphi_1^- & \varphi_2^{--} \end{pmatrix}\sim \left(1,\textbf{2},\textbf{2},-1\right)
\end{align}
that couples to the SM and mirror leptons that are doublets under the two different $\text{SU}(2)$ groups via the interaction $Y_\nu\; l \Phi^\dagger l'$. The scalar potential contains a trilinear term involving all Higgs multiplets $\kappa H \Phi^\dagger H'$ (see \eqref{eq:pot222} and \eqref{eq:VHPhi} for the potential) that induces a small vev far below the electroweak scale
\begin{align}\label{eq:vevPhi}
    v_\Phi \simeq -\frac{\kappa\;v\;v'}{\sqrt{2}\mu_\Phi^2}\ll v,
\end{align}
where $\mu_\Phi\gg v, v'$ is the bare mass of $\Phi$ and $v\;(v')$ the vev of $H\;(H')$. In this context we introduced the dimensionful parameter $\kappa<0$. The neutrino mass is given by $m_\nu = Y_\nu v_\Phi/\sqrt{2}$ without the need for small $Y_\nu$ and figure \ref{fig:numass} depicts a diagrammatic representation of this mechanism. After introducing the discrete symmetry and the particle spectrum in section \ref{sec:model}, we deal with the vacuum structure and the bosonic masses in sections \ref{sec:vacuum} and \ref{sec:bos}. The  strong CP problem is the focus of section \ref{sec:strong} and we demonstrate that loop corrections from the bidoublet do not spoil the symmetry-based solution to this problem. Following a brief discussion of low-energy constraints in \ref{sec:pheno} we consider the cosmological implications of our setup, especially for Dirac leptogenesis in \ref{sec:cosmo}.  The entire scalar potential and its minimization can be found in appendices \ref{sec:AppA}-\ref{sec:minima2} together with the sufficient conditions for vacuum stability in appendix \ref{sec:AppB}. Appendix \ref{sec:AppC} gives some additional details about the effective operator needed for Affleck-Dine leptogenesis.

\section{The model }\label{sec:model}
\begin{table}[t]
\centering
 \begin{tabular}{|c||c|c|c|c|c||c|} 
 \hline
  field&   $\text{SU(3)}_\text{C}$ & $\text{SU}(2)_\text{L}$ & $\text{SU}(2)'$ & $\text{U}(1)_\text{X}$&$\mathcal{Z}_3$  & $\text{generations}$\\
 \hline
 \hline 
 \addlinespace[0.3ex]
    $l$&  1 & $\textbf{2}$& $\textbf{1}$& $-1/2$ & $\omega^2$&  3\\
    $\overline{e}$ & 1 &  $\textbf{1}$ &  $\textbf{1}$ & 1& $\omega$&3\\
    \hline
    $q$ & 3 &  $\textbf{2}$&  $\textbf{1}$& $1/6$&$-$& 3\\
    $\overline{u}$ & $\overline{3}$ &  $\textbf{1}$ &  $\textbf{1}$ & $-2/3$&$-$&3\\
     $\overline{d}$ & $\overline{3}$ &  $\textbf{1}$ & $\textbf{1}$ & $1/3$&$-$&3\\
    \hline
    $H$& 1&  $\textbf{2}$&  $\textbf{1}$& $-1/2$&$-$&  1 \\
 \hline
  \hline
    $\Phi$ & 1 &  $\textbf{2}$ & $\textbf{2}$& $-1$&$-$& 1\\
\hline    
\hline
     $l'$&  1 &  $\textbf{1}$&  $\textbf{2}$& $-1/2$ &  $\omega$&3\\
    $\overline{e}'$ & 1 & $\textbf{1}$& $\textbf{1}$ & 1 &$\omega^2$&3\\
    \hline
    \addlinespace[0.3ex]
    $q'$ & $\overline{3}$ &  $\textbf{1}$&  $\textbf{2}$& $1/6$&$-$& 3\\
    $\overline{u'}$ & 3 &  $\textbf{1}$ &  $\textbf{1}$ & $-2/3$&$-$&3\\
     $\overline{d'}$ & 3 & $\textbf{1}$ & $\textbf{1}$ & $1/3$&$-$&3\\
    \hline
    $H'$& 1& $\textbf{1}$&  $\textbf{2}$& $-1/2$&$-$&  1 \\
 \hline
\end{tabular}
\caption{Charges and representations for the SM and mirror sector fields as well as the bidoublet $\Phi$. All spinors are left-chiral and we use the notation of \cite{Dreiner:2008tw}. We use $\omega \equiv e^{\frac{2\pi i}{3}}$. }
\label{tab:charges-reps}
\end{table}

\noindent This solution to the strong CP problem  hinges on a discrete exchange symmetry called \textit{generalized Parity} \cite{Craig:2020bnv} or \textit{Higgs-Parity}  \cite{Hall:2018let,Dunsky:2019api} that acts on the SM and mirror sector (indicated by a prime) via
\begin{align}\label{eq:exch}
    \psi(t,\vec{x}) \rightarrow i \sigma_2\; \psi'^*(t,-\vec{x}),
\end{align}
where $\sigma_2$ is the second Pauli matrix contracted with the spinor index, which is absent for the bosonic fields. Note that this symmetry is distinct from the typical exchange symmetry between both sectors, as every SM matter field is interchanged with the CP-conjugate of the corresponding mirror field  \cite{Chacko:2005pe}. The symmetry exchanges the $\text{SU(2)}$ gauge fields with each other.  For the gluon and  the $X$ boson the symmetry just acts on the spacetime arguments.  $\Phi$ gets mapped to its own CP-conjugate. The transformation properties of all fields are summarized in table \ref{tab:charges-reps}.
Up to the inclusion of the bidoublet this setup corresponds to model C of  \cite{Hall:2018let}. We use the two-component spinor formalism of \cite{Dreiner:2008tw}\footnote{Arrows on fermion lines denote the chirality structure and not the flow of fermion number. A bar is part of the particle label and does not denote any kind of conjugation.}.
$\text{U}(1)_\text{X}$ acts like hypercharge on all $\text{SU}(2)_\text{L}$ multiplets and $\text{SU}(2)_\text{L}\otimes \text{SU}(2)'$ singlets. Compared to the usual LRSM the abelian charges of all mirror doublets have the opposite sign, which is why we can not identify   $\text{U}(1)_\text{X}$   with $\text{U}(1)_\text{B-L}$.  It is worth pointing out that an $\text{SU}(2)_\text{L} \otimes \text{SU}(2)'$ bidoublet has exactly the same electric charge matrix as an $\text{SU}(2)_\text{L}$ triplet, so by charging it under $\text{U}(1)_\text{X}$  we obtain the desired electrically doubly charged component. The fermion sector (up to hermitian conjugates) is given by
\begin{align}
    \mathcal{L}_q &= Y_u q H^\dagger \overline{u} + Y_d q H \overline{d} + Y_u' q' H'^\dagger \overline{u}' + Y_d' q' H' \overline{d}',\label{eq:quarks}\\
    \mathcal{L}_l &= Y_l l H \overline{e} + Y_l' l' H' \overline{e}',\\
    \mathcal{L}_\text{port.}&= Y_\nu l\Phi^\dagger l'. \label{eq:port}
\end{align}
The SM (mirror) quarks and charged leptons obtain their masses solely from the vev of $H\;(H')$.
Non-observation of new colored fermion enforces a mirror scale of $v'\gtrsim 10^8\;\text{GeV}$ \cite{Craig:2020bnv,CMS:2018zkf,ATLAS:2018ziw}.
As a consequence of the $\text{U}(1)_\text{X}$ charge assignment there is no coupling between SM and mirror quarks and no coupling of  the bidoublet to any kind of quarks. We do not add electrically neutral singlet fermions. In the lepton sector there  would in principle exist three portal operators. The first one is the aforementioned  coupling to $\Phi$ displayed in \eqref{eq:port} and the other two $ \lambda_e l H \overline{e}'$  and $\lambda_e' l' H' \overline{e}$ would mix the electrically charged SM and mirror leptons. In this study we want to focus on the Type II Dirac Seesaw portal from $\Phi$, which is why we assume a lepton-specific $\mathcal{Z}_3$ symmetry (see the table \ref{tab:charges-reps}) under which $l, \overline{e'}$ transform as $\omega^2$ and  $l', \overline{e}$ as $\omega$ with $\omega \equiv e^{\frac{2\pi i}{3}}$ that removes the terms $\propto \lambda_e,\lambda_e'$. We can estimate the masses of the Dirac neutrinos from the vev \eqref{eq:vevPhi} to be
\begin{align}\label{eq:estimate}
   \frac{ m_\nu}{\SI{0.1}{\electronvolt}}  \simeq Y_\nu  \left(\frac{|\kappa|}{\SI{1}{\giga\electronvolt}}\right) \left(\frac{v'}{10^9\;\SI{}{\giga\electronvolt}}\right) \left(\frac{\SI{5e10}{\giga\electronvolt}}{\mu_\Phi}\right)^2.
\end{align}
To avoid large loop corrections to the bidoublet mass we will take $|\kappa| \ll \mu_\Phi$.
We choose a small value for $|\kappa|$  following the cosmological requirement  for baryogenesis in equation  \eqref{eq:kapp} of section \ref{sec:cosmo}. Setting $\kappa\rightarrow 0$ enhances the symmetry of the scalar potential \cite{Diaz:1997xv,Bonilla:2016zef}, since without the trilinear term in the potential \eqref{eq:VHPhi} we can rephase each multiplet independently, which is why a small value for $|\kappa|$ is technically natural \cite{tHooft:1979rat}. Owing to the fact that $v'\gg v$, for fixed $\mu_\Phi$ we have to take a smaller $|\kappa|$ than for the Majorana Type II Seesaw in order to have a sufficiently light $v_\Phi$.  $|\kappa|$ could have a dynamical origin via the vev of an additional scalar \cite{PhysRevD.25.774,Bonilla:2016zef} and a small vev could come from another iteration of the Type II Seesaw \cite{Grimus:2009mm,Gu:2009hu,Gu:2019yvw,Gu:2019ogb,Gu:2019ird}. Such a \textit{nested Seesaw} could arise schematically from a quartic term $\phi_1 \phi_2^3$ for two additional SM gauge singlet scalars $\phi_1$ and $\phi_2$, with $\phi_1$ being much heavier than the vev of $\phi_2$, resulting in its induced small vev $\braket{\phi_1}$  playing the role of $\kappa$. If we assume that   SM and mirror fermions have the same global B-L charge spectrum, then the mixed anomalies with the non-abelian gauge groups cancel separately for each sector. In this picture we see that the combined appearance of the terms $Y_\nu l \Phi^\dagger l'$ and $\kappa H \Phi^\dagger H'$ violates B-L by two units (as in Type II Seesaw models), because $H, H'$ are uncharged. The gauge symmetries and particle spectrum ensure that $l,l'$ do not pick up any Majorana mass terms allowed by $\Delta(\text{B-L})=2$, because they can only couple to each other via $\Phi$, but never to themselves in the absence of scalar $\text{SU}(2)_\text{L}$ and $\text{SU}(2)'$ triplets.

\section{Vacuum structure}\label{sec:vacuum}
\noindent 
The vev of the neutral component of $H'$ breaks $\text{SU}(2)'\otimes \text{U}(1)_\text{X}\rightarrow \text{U}(1)_\text{Y}$ and the discrete symmetry \eqref{eq:exch}, followed by the usual electroweak symmetry breaking induced by the vev of $H$. $\Phi$ contributes as a small perturbation to the spontaneous symmetry breaking (SSB) of all aforementioned symmetries due to its tiny vev.
We expand the multiplets into their components and assign vevs as 
\begin{align}\label{eq:ansatz}
    H \rightarrow \begin{pmatrix}  \frac{v}{\sqrt{2}}\\0\end{pmatrix}, \; H' \rightarrow  \begin{pmatrix}  \frac{v'}{\sqrt{2}}\\0\end{pmatrix}, \; \Phi \rightarrow  \begin{pmatrix}  \frac{ v_\Phi}{\sqrt{2}} & 0\\0 & 0 \end{pmatrix}.
\end{align}
There are two ways to generate the phenomenologically required hierarchy $v'\gg v$ between the mirror and SM Higgs vevs:
The first approach \cite{PhysRevD.41.1286} is to include soft breaking of the discrete exchange symmetry in the scalar potential $\mu_1^2 \left|H\right|^2 + \mu_2^2 \left|H'\right|^2$ with $\mu_1^2\ll\mu_2^2$.
It was shown recently \cite{deVries:2021pzl} that this soft breaking leads to two-loop contributions to the strong CP phase $\overline{\theta}$ (see section \ref{sec:strong}) in the original \textit{universal Seesaw} model by  \cite{PhysRevD.41.1286} regenerating the  $\overline{\theta}$ angle that was cancelled at tree-level. As of now there exists no similar analysis on the impact of soft breaking for the class of mirror sector models we are employing, so to be conservative we do not use this scheme.
A second mechanism was presented by \cite{Hall:2018let} that relies on tuning the quartic couplings of the scalar potential  in appendix \ref{sec:AppA} and the details will be discussed in appendix \ref{sec:minima}: If the mixed quartic coupling  $\lambda'$ in the potential
\begin{align}\label{eq:pot222}
    V &\supset  \lambda'   H^\dagger H\;H'^\dagger H' + \kappa \left( H \Phi^\dagger H' + H'^\dagger \Phi H^\dagger \right) 
\end{align}
(see \eqref{eq:VH} and  \eqref{eq:VHPhi} for the full potential) is set to zero, the scalar potential develops an unbroken custodial $\text{SU}(4)$ symmetry and one can view the   lighter SM-like Higgs as the  Goldstone-boson of this accidental symmetry. This idea is similar to the situation in the \textit{Twin-Higgs} model \cite{Chacko:2005pe}, where asymmetric vacua with $v\neq v'$ also require explicit breaking of an accidental custodial  $\text{SU}(4)$ \cite{Barbieri:2005ri}. However in those models the equivalent of the exchange symmetry \eqref{eq:exch} is typically softly broken as well \cite{Barbieri:2005ri}, whereas here the breaking is only spontaneous. The field $H'$ has a mass $-\mu_H^2<0$ and obtains the vev $v'\simeq \mu_H / \sqrt{\lambda_H}$.
After integrating out $H'$ one finds that the potential for  $H$ reads 
\begin{align}
     \lambda' v'^2   H^\dagger H   +  \lambda'\left(1+\frac{2\lambda'}{\lambda_H}\right) \left( H^\dagger H  \right)^2
\end{align}
and for spontaneous symmetry breaking one requires $\lambda'<0$.  Further $|\lambda'|\ll1$ is needed for the phenomenologically required hierarchy $v\ll v'$. The second term in the above is the self-coupling of $H$ modified by the finite threshold correction from integrating out $H'$  \cite{Elias-Miro:2012eoi}. In  \cite{Hall:2018let}   a real $v $ needs a small $\lambda'<0$ at the high scale $\mu =v'$, which is why in this construction, $v'$ is identified with the electroweak instability scale
\begin{equation}\label{eq:vprime}
    v'\simeq \left(10^9-10^{12}\right)\;\text{GeV}.
\end{equation}
RGE effects dominated by the top quark Yukawa then drive the Higgs self-coupling $\lambda_h$ to its positive $\mathcal{O}(0.1)$ value at low energies. Once we add a  bidoublet in \eqref{eq:pot222} and integrate it out we find that 
\begin{align}\label{eq:thr}
    \lambda_\text{eff}' \equiv \lambda' - \frac{\kappa^2}{\mu_\Phi^2}.
\end{align}
plays the role of $\lambda'$. The smallness and sign of this mixed quartic could be understood as the result of the threshold correction \cite{Elias-Miro:2012eoi} from $\Phi$, but since we actually have $\kappa^2/\mu_\Phi^2 \simeq  m_\nu \kappa /(Y_\nu  v v')\ll 1$  the correction to $\lambda'$ is completely negligible. As it turns out, the tree level potential is not enough for the correct vacuum structure and to induce $v\neq 0$ we actually need to include quantum corrections  \cite{Dunsky:2019api,Jung:2019fsp,Dunsky:2020dhn} from the one-loop Coleman-Weinberg potential  \cite{PhysRevD.7.1888}. This contribution, again dominated by the top quark, generates quartic scalar terms with a coupling $c_1<0$ (see section \ref{sec:minima} in the appendix for details). This results in a Higgs mass and self coupling of \cite{Jung:2019fsp}  
\begin{align}
m_h \simeq \sqrt{-\left(\lambda_\text{eff}'-\frac{c_1}{2}\right)}\;v',\quad  \lambda_h(\mu =v')=\frac{c_1}{16}\lesssim0 \label{eq:veq}.
\end{align}
A shortcoming of this approach is that one needs the fine-tuning $\lambda_\text{eff}'\simeq  c_1/2$ of order $\mathcal{O}(v^2/v'^2)$ for the  hierarchy $v'\gg v$ \cite{Hall:2018let} ($\lambda_\text{eff}'<0$ for a real $v$). In appendix \ref{sec:a} we demonstrate that the $v_\Phi$ in the Type II Seesaw regime does not spoil the desired vacuum structure. As far as naturalness is concerned, the small vev $v_\Phi$ in \eqref{eq:vevPhi} is technically natural \cite{tHooft:1979rat} and one may argue  along the lines of \cite{Chacko:2005pe,Hall:2018let}, that the hierarchy between $v$ and $v'$ does not lead to a separate hierarchy problem besides the usual one. Of course here will be loop corrections from the heavy $\Phi$, which could be cured by compositeness or supersymmetry at the bidoublet mass scale $\mu_\Phi > 10^{10}\;\text{GeV}$. If we consider the vevs $v_i$ to have phases $\beta_i$ then gauge transformations with the transformation parameters $\omega,\omega',\omega_X$ shift the phases to be\footnote{Note the slight abuse of notation for the passive gauge transformations.} \cite{Zhang:2007da}
\begin{align}
    \beta &\rightarrow \beta + \frac{1}{2}\left(\omega-\omega_X\right),\quad
     \beta' \rightarrow \beta' + \frac{1}{2}\left(\omega'-\omega_X\right),\\
      \beta_\Phi &\rightarrow \beta_\Phi + \frac{1}{2}\left(\omega+\omega'\right)-\omega_X.
\end{align}
If we set $\beta, \beta'$ locally to zero this induces the shift $\beta_\Phi \rightarrow \beta_\Phi-\beta-\beta'$ for the phase of $v_\Phi$. The minimization conditions of the scalar potential enforce that
\begin{align}\label{eq:phase-min}
   0= \frac{\partial V}{\partial \beta}=\frac{\partial V}{\partial \beta'}=-\frac{\partial V}{\partial \beta_\Phi} = \frac{\kappa\; v\; v'\; v_\Phi}{\sqrt{2}}\sin\left(\beta_\Phi-\beta-\beta'\right),
\end{align}
which implies that the physical phase for the vev $v_\Phi$ is zero. In other words, there is no spontaneous CP violation \cite{Lee:1974jb} in this model. The exchange symmetry \eqref{eq:exch} only enforces $Y_\nu = Y_\nu^\dagger$, so the PMNS Dirac phase would come from this  matrix if the charged lepton Yukawa $Y_l$ were purely real.

\section{Scalar and Gauge Bosons}\label{sec:bos}
\noindent The scalar spectrum consists of three CP even neutral scalars $h,h',h_\Phi = \sqrt{2}\;\text{Re}(\varphi_1^0)$ and one CP odd scalar  $a_\Phi =\sqrt{2}\; \text{Im}(\varphi_1^0)$  with the masses
\begin{align}
    m_h &\simeq \sqrt{2 \lambda_h}\; v,\quad
    m_{h'} \simeq  \sqrt{2 \lambda_H}\; v',\\
    m_{h_\Phi} &\simeq m_{a_\Phi}   \simeq \mu_\Phi,
\end{align}
where we used the low energy value $\lambda_h\simeq 0.129$ for the self-coupling of the SM like Higgs.
$h, h'$ mix primarily with each other via their quartic interaction and the small mixing angle is approximately $1/2 ( 1+  \lambda_\text{eff}' / \lambda_H) v/v'$.
The dominant source of mixing between $h\; (h')$ and $h_\Phi$ comes from the trilinear term and is $\simeq \kappa v'(v)/\mu_\Phi^2$. The same expression holds for the mixing between the \enquote{would-be-Nambu-Goldstone-bosons}(NGB)of the $Z \; (Z')$ gauge bosons (see the end of this section) with $a_\Phi$.
Additionally there are two singly-charged scalars $\varphi_1^\pm, \varphi_2^\pm$ and one doubly-charged scalar $\varphi_2^{\pm\pm}$ present. Their approximately degenerate masses read
\begin{equation}\label{eq:charged}
    m_{\varphi_1^\pm} \simeq m_{\varphi_2^\pm} \simeq m_{\varphi_2^{\pm\pm}} \simeq \mu_\Phi.
\end{equation}
There is also mixing between $\varphi_1^\pm \; (\varphi_2^\pm)$ and the \enquote{would-be-NGB} of the charged gauge bosons $W^\pm\;(W'^\pm)$ of the order of  $\simeq \kappa v'(v)/\mu_\Phi^2$.
When it comes to the charged gauge bosons we obtain
\begin{align}
    m_W = \frac{g}{2}\sqrt{v^2 + v_\Phi^2},\quad   m_{W'} = \frac{g}{2}\sqrt{v'^2 + v_\Phi^2}.
\end{align}
There is no mass mixing between the charged gauge bosons at tree-level since $\Phi$ only has one vev \cite{PhysRevD.18.1621}. Mixing could arise from loop diagrams involving the tree-level mixing between the electrically charged SM and mirror leptons in \eqref{eq:port}, which we have set to zero via another discrete symmetry. In the neutral gauge boson sector we find in addition to the massless photon that
\begin{align}
    m_Z \simeq \frac{g \sqrt{v^2+v_\Phi^2}}{2 \cos(\theta_W)},\quad m_{Z'} \simeq \frac{g}{2} \frac{\cos(\theta_W)^2}{\cos(2\theta_W)} v',
\end{align}
where we employed the weak mixing angle defined in \eqref{eq:weak}. 
It is evident that the bidoublet vev contributes with the same strength to $m_W$ and $m_Z$, which means that the SM prediction for the electroweak $\rho$-parameter \cite{ROSS1975135,Veltman:1976rt}  defined as $\rho \equiv m_W^2 /( m_Z^2 \cos(\theta_W)^2)$ is unchanged. To understand why, note that after the SSB of
$ \text{SU}(2)'\otimes \text{U}(1)_\text{X}$ down to $\text{U}(1)_\text{Y}$, the multiplet $\Phi$ decomposes into two $\text{SU}(2)_\text{L}$ doublets $\Phi_1\equiv \left(\varphi_1^0,\varphi_1^-\right)^t$ with $Y=-1/2$ and  $\Phi_2\equiv\left(\varphi_2^-,\varphi_2^{--}\right)^t$ with $Y=-3/2$. Since only the neutral component of $\Phi_1$ develops a vev the contribution of $\Phi$ to the SM gauge boson masses reduces to the one  in a Two-Higgs-doublet model. This is why at tree-level our model does not modify the $\rho$-parameter and  it can not help to address the tentative tension in the $W^\pm$ boson mass reported by the CDF collaboration \cite{CDF:2022hxs}. Moreover, unlike for the electroweak triplet needed for the conventional Type II Seesaw,  here the $\rho$-parameter does not force the small vev to be below the GeV-scale \cite{Kanemura:2012rs}. In principle, there could also be one-loop gauge boson self-energy diagrams with e.g.  $h_\Phi$ and $\varphi_1^\pm$ running in the loop \cite{Heeck:2022fvl}. The shift in the relevant electroweak precision observables \cite{PhysRevLett.65.964,PhysRevD.46.381} will roughly depend on their mass splitting via $(m_{h_\Phi}^2 -   m_{\varphi_1^\pm}^2)/\mu_\Phi^2$.
However, since we assume all mass splittings to be small compared to the largest scale in the scalar potential $\mu_\Phi^2$ and since the contribution will essentially decouple for large bidoublet masses, our model can not help ameliorate the CDF tension \cite{CDF:2022hxs}. 
When it comes to the mixing between the neutral gauge bosons the situation simplifies in the limit $v_\Phi\rightarrow 0$: There are only two mixing angles required. The electroweak mixing angle is defined via\footnote{The discrete exchange symmetry requires the $ \text{SU}(2)'\otimes  \text{SU}_\text{L}(2)$ couplings $g',g$ to be equal at high scales and we neglect the differences in their RGE running here.} \cite{PhysRevD.41.1286}
\begin{align}\label{eq:weak}
    \frac{g_X}{g} = \frac{\sin(\theta_W)}{\sqrt{\cos(2\theta_W)}}
\end{align}
and the angle between the physical  $Z, Z'$   \cite{PhysRevD.41.1286} reads
\begin{align}\label{eq:gamma}
    \sin(\gamma) = \frac{\sin(\theta_W)^2 \sqrt{\cos(2\theta_W)} }{\cos(\theta_W)^4}\left(\frac{v}{v'}\right)^2.
\end{align}
Turning on $v_\Phi$ only leads to a sub-dominant modifications of the angle $\gamma$ as long as $v_\Phi \ll v$.

\begin{figure*}[t]
 \centering
  \includegraphics{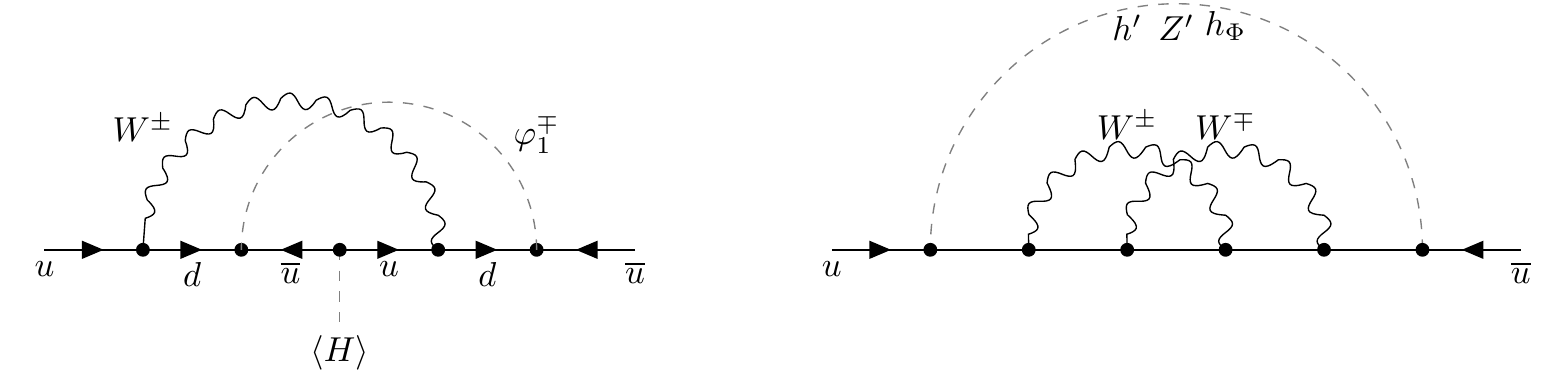}
  \caption{Two- $\textit{(left)}$ and three-loop $\textit{(right)}$ Feynman diagrams leading to phases in the quark mass matrices contributing to $\overline{\theta}=\theta_\text{chir.}$ in this model.
  $h', Z', h_\Phi$ and $\varphi_1^\pm$ only couple  to the SM quarks via suppressed mixing. For the three- loop diagrams we did not indicate the internal chirality structure and the labels of the internal quark fields, because for $h,h_\Phi$ there is a mass insertion $\propto \braket{H}$ after the first vertex and in the $Z'$ case there is the same kind of insertion before the sixth vertex.}
  \label{fig:cp}
\end{figure*}
\section{Strong CP problem}\label{sec:strong}

\noindent The physical CP violating parameter   $\overline{\theta}= \theta_\text{QCD}+\theta_\text{chir.}$  is conventionally split into the contribution of the QCD theta-term and the part $\theta_\text{chir.}= \text{arg}\left(\text{det}\left(M_u M_d\right)\right)$ arising from the up- and down-type quark mass matrices $M_u, M_d$. 
Following from the fact that the topological vacuum selection parameter $\theta_\text{QCD}$ arises because of non-perturbative QCD dynamics and due to its origin as the coefficient of a non-vanishing surface term \cite{Senjanovic:2020int}, one might argue that  $\theta_\text{QCD}$ is unlike all other dimensionless parameters of the SM such as gauge or Yukawa couplings and more akin to a boundary condition. In the SM, the electroweak part $\theta_\text{chir.}$ only receives finite loop corrections at three-loop order and diverges at seven loops \cite{Ellis:1978hq}, which is fundamentally different from e.g. the hierarchy problem of the Higgs mass. In recent years, a new perspective  on the strong CP problem has emerged \cite{Ai:2020ptm,Ai:2022htq} that relies on a careful analysis of the boundary condition for the path integral and the infinite spacetime volume limit, suggesting that the strong CP violation disappears for the mathematically correct order of limites. In the present work we take the smallness of $\overline{\theta}$ at face value and follow the UV symmetry based BSM approach \cite{Mohapatra:1978fy,Nelson:1983zb,Barr:1984qx,PhysRevD.41.1286,PhysRevLett.67.2765,Hook:2014cda} to \enquote{explain} its tiny value. For recent work that directly ties the smallness of $\overline{\theta}$ to a different Dirac neutrino mass generation mechanism see \cite{Carena:2019nnd}.
\subsection{Tree level}
\noindent 
Owing to the fact that the discrete exchange symmetry defined in \eqref{eq:exch} imposes $\theta_\text{QCD}=0$  we only need to care about the quark contribution. Following reference  \cite{Hall:2018let} there could exist a dimension six operator allowed by the discrete exchange symmetry in \eqref{eq:exch}
\begin{align}\label{eq:dim6}
    \frac{c_6}{\Lambda_\text{UV}^2} \left(H^\dagger H - H'^\dagger H'\right)G_{\mu\nu}  \tilde{G}^{\mu\nu}
\end{align}
that regenerates $\theta_\text{QCD}$ after the $H'$ obtains a vev. This leads to the requirement of $v'<10^{13}\;\text{GeV}$ for a cut-off scale of $\Lambda_\text{UV}=M_\text{Pl.}$ and order one Wilson coefficient $c_6$, to stay within the observational bound of $\overline{\theta}<10^{-10}$ \cite{PhysRevD.19.2227,CREWTHER1979123,Baker:2006ts,2016PhRvL.116p1601G}. For the given particle content this operator is not realized at the loop level. The mass matrix for either up-type or down-type  quarks in the basis $\left(q',\overline{q}\right)$ and $\left(q,\overline{q}'\right)^t$ with $q=u,d$, where we have suppressed generation indices, reads
\begin{equation}
    M_q = \begin{pmatrix} 0& Y_q' \frac{v'}{\sqrt{2}}  \\   Y_q \frac{v}{\sqrt{2}} &0\end{pmatrix}.
\end{equation}
Because the  exchange symmetry \eqref{eq:exch} sets $Y_q' = Y_q^*$ \cite{Hall:2018let}
 \begin{align}
    \text{arg}\left(\text{det}\left(M_q\right)\right)=-\frac{v v'}{2}\text{arg}\left(\text{det}\left(Y_q\right)\;\text{det}\left(Y_q'\right)\right)
\end{align}
vanishes, meaning that the SM and mirror sector phases cancel each other out \cite{PhysRevLett.67.2765}. Since neither $Y_q$ nor $Y_q'$ are required to be real,  they source the  CKM phase for the SM and mirror sector and one does not need a separate sector to do so unlike in the case for  Nelson-Barr models \cite{Nelson:1983zb,Barr:1984qx}.  The presence of the bidoublet field does not change this picture as it does not couple to quarks.  Reference \cite{Kawamura:2018kut} demonstrated that integrating out the heavy mirror quarks does not generate phases for the SM quark Yukawas via RGE effects.
\subsection{Loop level}
\noindent So far we have only worked at tree-level. Radiative corrections to the quark masses at one-loop level all turn out to be real valued. Two-loop diagrams with two $W^\pm$ running in the loops have the correct complex couplings from the CKM matrix but the wrong chirality structure \cite{Ellis:1978hq}. That leaves us with two options: Either we replace one of the $W^\pm$ with the charged scalar $\varphi_1^\pm$, that couples to quarks via its mixing with the \enquote{would-be-NGB} of the $W^\pm$, and add a mass insertion for the right-chirality structure (see the left diagram in figure \ref{fig:cp})
\begin{align}
  \theta^{(2)}\simeq  \frac{\alpha}{\pi} \left(\frac{\kappa v'}{\mu_\Phi^2}\right)^2 \frac{m_q^2}{m_W^2} \frac{m_q^2}{\mu_\Phi^2},
\end{align}
or we add a third loop with a neutral boson \cite{Ellis:1978hq} depicted on the right in figure \ref{fig:cp}
\begin{align}
  \theta^{(3)}\simeq  \left(\frac{\alpha}{\pi}\right)^3  \left(\frac{m_q^2}{m_W^2}\right)^3 \begin{cases} 
  \left(\frac{v}{v'}\right)^2 \frac{m_q^2}{m_{h'}^2}\; &(h'),\\
  \left(\frac{v}{v'}\right)^4\; \frac{m_q^2}{m_W^2} \frac{m_q^2}{m_{Z'}^2}\; &(Z'),\\
  \left(\frac{\kappa v'}{\mu_\Phi^2}\right)^2 \frac{m_q^2}{\mu_{\Phi}^2}\; &(h_\Phi).
  \end{cases} 
\end{align}

\noindent In the above estimates we have dropped order one and loop factors. $m_q^2$ must be a combination of two different quark masses due to CKM unitarity and $\alpha$ denotes the fine-structure constant. All of the above contributions are negligibly small due to large BSM mediator masses and small mixing angles. For instance the factor $\kappa v'/ \mu_\Phi^2\simeq m_\nu /(Y_\nu v)\simeq 10^{-12}$ for $Y_\nu = \mathcal{O}(1)$ appearing for the bidoublet scalars is already sufficient to suppress the loop diagrams below the current experimental bounds of $\overline{\theta}<10^{-10}$ \cite{PhysRevD.19.2227,CREWTHER1979123,Baker:2006ts,2016PhRvL.116p1601G} and the small ratio $v/v'$ achieves the same.  Leptonic loops involving the $Y_\nu$ coupling occur at even higher orders and are even more negligible. 
Consequently we find that the leading contribution arises from purely SM effects at three loops (two virtual $W^\pm$  and one virtual gluon) and reads $\overline{\theta}\simeq \mathcal{O}\left(10^{-16}\right)$ \cite{Ellis:1978hq} which corresponds to an electric dipole moment of the neutron of about $\mathcal{O}\left(10^{-31}\right)\;e\;\text{cm}$  \cite{Ellis:1978hq}.
Unfortunately this is still out of reach for current and future experiments, that are expected to probe dipole moments down to $\mathcal{O}\left(10^{-27}\right)\;e\;\text{cm}$ \cite{Lamoreaux:2009zz,Baker:2011xit,Tsentalovich:2014mfa}.

\begin{figure*}[t]
 \centering
  \includegraphics{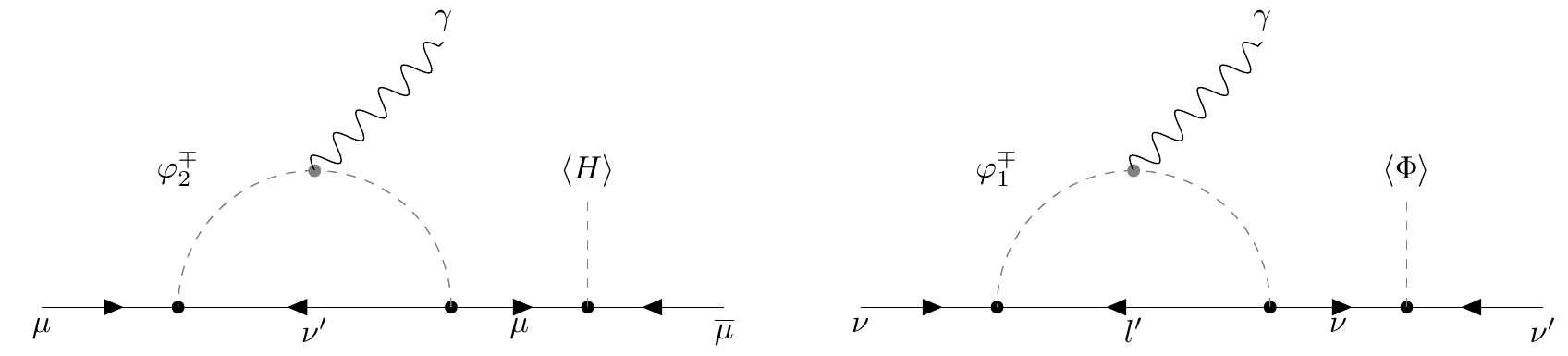}
  \caption{One-loop Feynman diagrams contributing to the  magnetic moments of the muon $\textit{(left)}$ and neutrinos $\textit{(right)}$. The photon line in the second diagram can be attached to the electrically charged mirror lepton inside the loop as well. For this diagram the mass insertion can also appear on the incoming line, so that we get a second set of diagrams with $l,\varphi_2^\mp$ running in the loop.}
  \label{fig:mag-mom}
\end{figure*}
\section{Low-energy phenomenology}\label{sec:pheno}

\noindent Tree-level exchange of $\varphi_2^-$ leads to a BSM contribution to muon decay of \cite{Babu:2019mfe}
\begin{align}
    \Gamma (\mu^- \rightarrow \sum_{i,j} e^- \nu_i'^\dagger \nu_j' )\simeq \frac{1}{6144\pi^3}\frac{m_\mu^5}{\mu_\Phi^4} \sum_{i,j} \left|\left(Y_\nu\right)_{\mu i} \left(Y_\nu\right)^*_{e j}\right|^2,
\end{align}
which modifies the Michel-parameters $(\rho, \delta, \xi)$ \cite{Michel:1949qe,PhysRev.106.170} encoding the angular and energy distribution of the decay relative to the SM. Using the methods of \cite{Kuno:1999jp} we find that $\rho = \xi \delta = 3/4 \delta = 3/16 |g_S|^2$ when compared to the SM where the number on the right hand side is one. In this context we have defined
\begin{equation}
    g_S \equiv \frac{v^2}{2\mu_\Phi^2} \sum_{i,j} \left(Y_\nu\right)_{\mu i} \left(Y_\nu\right)^*_{\mu j},
\end{equation}
which is currently constrained to be smaller than 0.55 \cite{MISHRA1990170,CHARM-II:1994dzw} not imposing any stringent limits on our scenario with super-heavy scalars. Note that the previous bound was derived using left-chiral neutrinos scattering off charged leptons \cite{CHARM-II:1993phx,CHARM-II:1994dzw,TEXONO:2009knm,Beda:2009kx,Beda:2010hk}, whereas our decay involves the $\nu'$ of opposite chirality, so we expect that the limit on $g_S$ in our model would be even weaker due to additional neutrino mass insertions.
\subsection{Dipole moments and lepton flavor violation}
\noindent Loops involving $\varphi_2^\pm$ generate a correction to the magnetic dipole moment of the muon depicted in the left diagram of figure \ref{fig:mag-mom}
of 
\begin{align}
   \Delta a_\mu \simeq \frac{e}{96\pi}  \left(\frac{m_\mu}{\mu_\Phi}\right)^2 \;\sum_{j=e,\mu,\tau} \left(Y_\nu\right)_{\mu j} \left(Y_\nu\right)^*_{\mu j}.
\end{align}
Here, there is no chiral enhancement inside the loop and the correct chirality structure is obtained from a mass insertion on the external legs (see figure \ref{fig:mag-mom}), hence the dependence on $m_\mu$.
For masses of $\mu_\Phi \simeq 10^{10}\;\text{GeV}$ (see \eqref{eq:estimate}), the shift in the magnetic moment is of $\mathcal{O}\left(10^{-36}\right)$, which is far too small to explain the deviation of $\Delta a_\mu =(251\pm 59)\times 10^{-11}$ observed by the BNL \cite{Muong-2:2006rrc} and FNL \cite{Muong-2:2021ojo} collaborations.
We can reuse this result to estimate the full transition dipole form factor and find the partial width \cite{Lavoura:2003xp}
\begin{align}
    \frac{\text{BR}\left(\mu\rightarrow e\gamma\right)}{8\times10^{-8}} \simeq  \alpha  \left(\frac{v}{\mu_\Phi}\right)^4
  \left|\sum_{j=e,\mu,\tau} \left(Y_\nu\right)_{\mu j} \left(Y_\nu\right)^*_{e j} \right|^2.
\end{align}
Compared to the present experimental limit of $\text{BR}\left(\mu\rightarrow e\gamma\right)<4.2\times10^{-13}$ \cite{MEG:2016leq} set by the MEG collaboration and the future projection of $6\times10^{-14}$ from MEG II \cite{MEGII:2018kmf},  our scenario leads to branching ratios of $\mathcal{O}(10^{-40})$ for the $\mu_\Phi$ in \eqref{eq:estimate} and is therefore not excluded.
Since the bidoublet only connects leptons and mirror leptons, the process $\mu^-\rightarrow e^- e^+ e^-$  occurs via a penguin diagram with the same dipole form factor as before so we can estimate  $ \text{BR}\left(\mu^-\rightarrow e^- e^+ e^-\right)\simeq 7\times 10^{-3}\; \text{BR}\left(\mu\rightarrow e\gamma\right)$ \cite{Hisano:1998fj}, which is compatible with the current  bound $ \text{BR}\left(\mu^-\rightarrow e^- e^+ e^-\right)<10^{-12}$ from \cite{BELLGARDT19881} and the projected sensitivity of $\mathcal{O}(10^{-15})$ of the Mu3e experiment \cite{Perrevoort:2018ttp}. The analogous decays of $\tau$ leptons are typically less constrained and also do not set any significant bounds on our scenario. Similarly we can estimate the neutrino magnetic moment, where there are two diagrams involving the coupling of $\varphi_1^\pm\; (\varphi_2^\pm)$ to $\nu\;(\nu')$ depicted on the right side of figure \ref{fig:mag-mom}:
\begin{align}
    \left(\mu_\nu\right)_{ii} = \frac{\mu_B}{16\pi^2} \frac{m_e (m_\nu)_i}{\mu_\Phi^2}  \sum_{j=e,\mu,\tau} \left(Y_\nu\right)_{i j} \left(Y_\nu\right)^*_{i j}
\end{align}
Both diagrams contribute with the same strength as $\varphi^\pm_1$ and  $\varphi^\pm_2$ are mass degenerate, see \eqref{eq:charged}. Here the factor of $m_e$ does not arise from any chirality enhancement but rather from the definition of the Bohr magneton $\mu_B \equiv e/(2 m_e)$. Again we observe mass insertions on the external legs (see figure \ref{fig:mag-mom}) for the right-chirality structure explaining the $m_\nu$ dependence. The most stringent limit on neutrino magnetic moments of $\mu_\nu < 6.3\times 10^{-12}\;\mu_B$ comes from the XENONnT experiment \cite{XENON:2022mpc} and our estimate for the aforementioned masses reads  $\mu_\nu \simeq \mathcal{O}(10^{-36})\;\mu_B$,  far below the bound.
\subsection{Collider bounds}
\noindent The singly-charged scalars $\varphi_{1,2}^\pm$ have to be heavier than $\mathcal{O}(\SI{100}{\giga\electronvolt})$  to escape direct production at colliders \cite{ALEPH:2001oot,OPAL:2003nhx,L3:2003fyi,DELPHI:2003uqw}.
If we were to turn on the couplings $\propto \lambda_e, \lambda_e'$ between the charged SM and mirror leptons   the  $\varphi_2^{\pm\pm}$ could produce same-sign di-lepton signatures, similar to the canonical Type II Seesaw. Current collider searches \cite{ATLAS:2017xqs} place a bound of $m_{\varphi_2^{\pm\pm}}> \SI{800}{\giga\electronvolt}$ and a future $\SI{100}{\tera\electronvolt}$ proton-proton collider could probe masses up to  $\SI{4.5}{\tera\electronvolt}$ \cite{Du:2018eaw}. In this scenario, the exchange of the neutral $h_\Phi$ can induce a contact interaction between the SM leptons, which evades the LEP bound \cite{Electroweak:2003ram} due to the large $\mu_\Phi$ and potentially small mixing between SM and mirror leptons. For the smallest allowed $v'\simeq 10^9\;\text{GeV}$, we find that the mirror electron (see the discussion above \eqref{eq:TRH} in the next section) would have a mass of $\simeq\SI{2}{\tera\electronvolt}$, potentially accessible at colliders.

\section{Cosmology}\label{sec:cosmo}
\subsection{Reheating}
\noindent In the early universe the  discrete exchange symmetry in \eqref{eq:exch} is spontaneously broken by the vev of the heavy doublet $H'$, leading to the presence of topological defects, which can overclose the universe if they are stable \cite{Kibble_1976,1975JETP...40....1Z,PhysRevLett.48.1156}. There exist basically two remedies for this conundrum: One may either include small bias terms \cite{PhysRevLett.48.1156,PhysRevD.39.1558} in the scalar potential, explicitly breaking the discrete symmetry  and thereby leading to domain wall decay. The explicit breaking might then manifest \cite{deVries:2021pzl} as a contribution to $\overline{\theta}$ at low energies  similar to the soft breaking discussed in section \ref{sec:vacuum}. 
Alternatively \cite{Guth:1980zm}, if the domain walls are formed before or during  the   exponential expansion phase of cosmic inflation, they will be diluted by the expansion of spacetime.
The second scenario requires that the symmetry is broken before or during inflation and does not get restored afterwards, which can be satisfied for a reheating temperature of $T_\text{RH}<v'$ \cite{DAgnolo:2015uqq}. Since  the discrete exchange symmetry  relates the SM and mirror Yukawas, we expect a similar mass spectrum in the mirror sector up to factors of $v'/v$ of course. This means that, as long as we turn of the lepton-mirror lepton mixing in \eqref{eq:port} via the $\mathcal{Z}_3$ symmetry in table \ref{tab:charges-reps},  the mirror electron $e'$ is the lightest  stable electrically charged particle of the mirror sector. To avoid the stringent bounds \cite{Yamagata:1993jq,PhysRevLett.68.1116,PhysRevD.41.2074,Norman:1986ux} on the number density of such charged thermal relics \cite{Berger:2008ti}, we require the reheating temperature to be
\begin{align}\label{eq:TRH}
    T_\text{RH} < m_{e'} \simeq 2\times 10^{-6}\cdot v'\;,
\end{align}
corresponding to  $ T_\text{RH} < 2\times(10^3-10^6)\;\text{GeV}$ for $v'=(10^9-10^{12})\;\text{GeV}$.
Of course there are also mirror quarks, with the lightest quark having a mass of $m_{u'}=m_u v'/v \simeq 2\times 10^{-5} v'$ one order of magnitude above \eqref{eq:TRH}.
Reference \cite{Dunsky:2019api} found that the mirror quark masses actually  run faster compared to the mirror leptons owing to their color charge, leading to a situation where the lightest mirror quarks $u',d'$ are almost mass degenerate with $e'$ for $v'\gg 10^{11}\;\text{GeV}$. Consequently relic abundances of colored mirror fermions are also avoided by the previously determined reheating temperature. Reheating could either occur from the dynamics of the oscillating inflaton condensate or from a second unrelated epoch of intermediate matter domination \cite{Scherrer:1984fd}.
Alternatively one might consider asymmetric reheating scenarios \cite{Craig:2016lyx,Chacko:2016hvu}, in which the SM and mirror sectors  are reheated to different temperatures. This could happen if the particle responsible for reheating decays preferentially to the SM instead of the mirror sector.
As a consequence of the large hierarchy between $v$ and $v'$, the mirror neutrinos never equilibrate with the SM plasma via gauge or Yukawa interactions and are only produced via freeze-in \cite{Luo:2020sho,Luo:2020fdt}.
\subsection{Dark Radiation}
\noindent Since the present setup   only doubles the $\text{SU}(2)$ gauge group of the SM, without introducing a second $\text{U}(1)$, there is no dark photon. Thus the associated problem of large amounts of dark radiation from mirror neutrinos and a dark photon, that typically plaques mirror sector models \cite{Craig:2016lyx}, is absent.  For $2\rightarrow 2$ scattering producing $\nu'$ from the SM,  $h_\Phi$ exchange is completely negligible due to its large mass. $Z$ exchange via $Z-Z'$ mixing (see \eqref{eq:gamma}) leads to
\begin{align}
    \frac{\Delta N_\text{eff.}}{ \mathcal{O}(10^{-14})} \simeq  \left(\frac{10^9\;\text{GeV}}{v'}\right)^4
    \begin{cases}
    \left(\frac{T_\text{RH}}{\SI{100}{\giga\electronvolt}}\right)^3\; &(T_\text{RH}\lesssim v),\\
   0.1 \cdot\left(\frac{\SI{1}{\tera\electronvolt}}{T_\text{RH}}\right)\; &(T_\text{RH}\gg v),
    \end{cases}
\end{align}
and we find that out-of-equilibrium $Z$ decays to two $\nu'$ would give  $\Delta N_\text{eff.}\simeq   10^{-15} \left(10^9\;\text{GeV}/ v'\right)^4$.
These yields are at least two orders of magnitude smaller than the contribution $\Delta N_\text{eff.}\simeq 7.5\times 10^{-12}$ from out-of-equilibrium Higgs decays \cite{Luo:2020fdt}, provided that $T_\text{RH}\gtrsim m_h$, and Higgs mediated scattering leads to $\Delta N_\text{eff.}<10^{-10}$. 
\subsection{Leptogenesis from decays}
\noindent Seesaw mechanisms are often invoked to realize baryogenesis via the leptogenesis mechanism \cite{FUKUGITA198645}. For the standard out-of-equilibrium decay  scenario it has long been known that scalar triplet leptogenesis  \cite{Hambye:2005tk} requires at least two triplets or insertions of heavy neutrinos to generate the required CP violation. Otherwise there would be no imaginary part in the interference term between the tree-level decay and its one-loop self-energy and vertex corrections. This conclusion also holds for Dirac Seesaw models \cite{Gu:2006dc,Gu:2012fg} and we could consider the channel  $\Phi\rightarrow l l'$, which also requires  $\Phi\rightarrow H H'$ for the asymmetry generation via self-energy graphs \cite{Gu:2006dc,Gu:2012fg} for at least two different bidoublets. The tree level decay widths of each bidoublet read (with suppressed generation indices)
\begin{align}\label{eq:modes}
    \Gamma(\Phi\rightarrow l l ')= \frac{ Y_\nu^2}{8\pi} \mu_\Phi, \quad    \Gamma(\Phi\rightarrow H H ')=  \frac{\kappa^2}{ 32\pi  \mu_\Phi}.
\end{align}
We emphasize that the low reheating temperature in \eqref{eq:TRH} is in tension with the high scale $>\mathcal{O}(10^{10}\;\text{GeV})$ bidoublet mass, which is why non-thermal leptogenesis \cite{Lazarides:1990huy} would be required. 
As an example we consider reheating via  perturbative decays of an inflaton with mass $m_I>2\mu_\Phi$, decaying to both SM particles and bidoublets. The inflaton's total decay width $\Gamma_I$ is related to the  reheating temperature via $T_\text{RH}\sim \sqrt{\Gamma_I M_\text{Pl.}}$. One finds that the baryon asymmetry normalized to entropy at the end of reheating would be given by \cite{Asaka:2002zu}
\begin{equation}\label{eq:NT}
    \frac{n_B}{s}\simeq -\frac{42}{79}\;\varepsilon\; \text{BR}_I\; \frac{T_\text{RH}}{m_I}.
\end{equation}
Here $\text{BR}_I<1$ denotes the branching ratio of inflaton decays to bidoublets and $\varepsilon$  is the previously mentioned CP violating decay parameter depending on the mass spectrum of the different bidoublet generations \cite{Gu:2012fg}. As for all Dirac leptogenesis scenarios \cite{Dick:1999je}, equal and opposite asymmetries in $l$ and $l'$ are produced. If the $\text{SU}(2)'$ sphalerons are fast during or after the asymmetry generation, then the asymmetry in $l'$ will be transferred into a mirror baryon asymmetry, which will be equal and opposite to the baryon asymmetry produced via $\text{SU}(2)_\text{L}$ sphalerons from the $l$ asymmetry. Since there is no direct interaction coupling baryons to mirror baryons, the respective asymmetries will not equilibrate to zero and remain separately conserved.
For a hierarchical bidoublet spectrum with the lightest mass $\mu_\Phi$ one finds that the CP violating decay parameter reads  \cite{Hambye:2005tk, Gu:2012fg}
\begin{align}
    \varepsilon< \frac{ r\; \sqrt{\text{BR}_l\; \text{BR}_H}}{8\pi}  \frac{m_\nu\;\mu_\Phi}{ v\; v'} ,
\end{align}
where $r\equiv \mu_\Phi / \mu^{(2)}_{\Phi}<1$ is the ratio of the lightest and next heavier bidoublet masses, $\text{BR}_{l,H}$ are the branching ratios for both decay modes in \eqref{eq:modes} and $m_\nu$ is the heaviest active neutrino mass. A typical value for equal branching fractions and $\mu_\Phi = 10\; v'$ is $\varepsilon\simeq 10^{-13}$, which is smaller than for the Type II Seesaw result \cite{Hambye:2005tk} due to the additional $v/v'$ suppression. For a hierarchical bidoublet spectrum we find  
\begin{align}
   \left| \frac{n_B}{s}\right|^\text{hier.} < 10^{-23} \left(\frac{\text{BR}_I}{1\%}\right) \left(\frac{m_\nu}{ 0.1 \text{eV}}\right) \left(\frac{2 \mu_\Phi\; r }{ m_I}\right) \left(\frac{T_\text{RH}}{10^{-6}\;v'}\right)
\end{align}
being far too small to explain the observed value of $n_B/s \simeq 8\times10^{-11}$ \cite{Davidson:2008bu}. Therefore we have to invoke a resonant enhancement of the self-energy diagrams \cite{Liu:1993tg,Flanz:1994yx,Flanz:1996fb,Covi:1996wh,Pilaftsis:1997dr,Pilaftsis:1998pd,Pilaftsis:2003gt} via assuming that $\mu_\Phi$ and the next heavier mass $\mu_{\Phi}^{(2)}$ are nearly degenerate $|\mu_{\Phi}^{(2)}-\mu_\Phi|\ll \mu_{\Phi}^{(2)}\simeq \mu_\Phi  \quad (r\simeq 1)$. This scenario enhances the previous estimate for $\varepsilon$   by a factor of  \cite{Pilaftsis:1998pd,Pilaftsis:2003gt}
\begin{align}
    \frac{\rho }{\rho^2+\delta^2}\gg1,\quad \text{with} \quad \rho\equiv 1- r^2,\quad \delta \equiv \frac{\Gamma_\text{tot.}}{\mu_\Phi}.
\end{align}
The above expression is regulated by  the total decay width of the bidoublet   $\Gamma_\text{tot.}$, which is the sum of the rates in \eqref{eq:modes}. We assume both bidoublets to have comparable decay rates. A sizeable enhancement requires $\rho\sim\delta \ll 1$ and hence $ Y_\nu^2 + \kappa^2 /(4 \mu_\Phi^2)\ll 8\pi$. The decay width to Higgses is automatically small for $\kappa\ll \mu_\Phi$, but we may have to make $Y_\nu$ small by hand, which would require a larger $v_\Phi$  for this scenario to fit $m_\nu$. Note that this small $\Gamma_\text{tot.}\ll\mu_\Phi$ does not necessarily force $\varepsilon$ to be small, as this parameter depends only on the branching ratios  $\text{BR}_l \;\text{BR}_H = \Gamma(\Phi\rightarrow l l ')\Gamma(\Phi\rightarrow H H ')/\Gamma_\text{tot.}^2\leq 1/4$ and not on the absolute widths. Of course we can not make the decay width arbitrarily small, or else the decay will take place after inflationary reheating during an epoch where the bidoublets dominate the energy density of the universe. In this regime \eqref{eq:NT} still holds with the replacement $T_\text{RH}/m_I\rightarrow T_\text{dec.}/\mu_\Phi$ \cite{Giudice:1999fb}, where $T_\text{dec.}$ is the reheating temperature after the second matter dominated epoch. The enhancement factor of $\varepsilon$ is bounded from above by the perturbativity requirement $\varepsilon\ll 1$ assumed in the derivation of the Boltzmann equations, where one linearizes in the chemical potentials \cite{Giudice:2003jh,Davidson:2008bu}. The precise value of $\varepsilon$ depends on the details of the active neutrino mass spectrum such as almost degenerate masses \cite{Hambye:2003rt}, which is why we use $\varepsilon$ as a free parameter. 
Employing the kinematic condition $m_I > 2 \mu_\Phi$ and  \eqref{eq:TRH} to eliminate $T_\text{RH}/m_I$ in \eqref{eq:NT} lets us determine that there is indeed a parameter range reproducing the observed baryon asymmetry
\begin{align}
    \left|\frac{n_B}{s}\right|^\text{res.}<10^{-10}\left(\frac{\varepsilon}{0.05}\right)\left(\frac{\text{BR}_I}{5\%}\right)\left(\frac{50}{\mu_\Phi/v'}\right).
\end{align}
To obtain this result we had to set $\varepsilon$ close to its perturbative limit, which implies highly degenerate bidoublets with $|\mu_{\Phi}^{(2)}-\mu_\Phi| / \mu_\Phi\simeq 10^{-12}$. We further had to assume only a small hierarchy between $v'$ and $\mu_\Phi$ to accommodate the inflaton decaying mostly to other SM particles implying $\text{BR}_I\ll1$. 
\subsection{Inflationary Affleck-Dine Leptogenesis}
\noindent Alternatively, the coherent rotation in field space of a complex scalar field with lepton number during inflation facilitates leptogenesis via the Affleck-Dine mechanism \cite{AFFLECK1985361}.
In this picture, the Sakharov conditions \cite{Sakharov_1991} are realized via the initial phase of the scalar  field providing C and CP violation and deviations from thermal equilibrium appear in the form of a large field amplitude during cosmic inflation. The last ingredient is baryon number violation, that arises from lepton number violation in the scalar potential and gets transmitted to the SM leptons so that afterwards it gets converted into baryon number  via the B+L violating $\text{SU}(2)_\text{L}$ sphaleron vertex. The authors of \cite{Barrie:2021mwi,Barrie:2022cub} put forth a very economical framework unifying Higgs inflation \cite{Bezrukov:2014bra} and the conventional Type II Seesaw. 
Motivated by \cite{Barrie:2021mwi,Barrie:2022cub}, we will assume that the inflaton is a linear combination of the neutral fields $h, h'$ and $h_\Phi+i a_\Phi\equiv\rho_\Phi e^{i\phi }$ after giving all scalar multiplets a non-minimal coupling to gravity \cite{STAROBINSKY198099}. 
Note that identifying the Affleck-Dine field with the inflaton is just a particularly convenient example for generating the required large initial field value and there exist other scenarios \cite{Dine:1995uk,Co:2020dya}, where the large field value is dynamically realized without this identification. Following the discussion at the end of section \ref{sec:model}, we  assign the B-L charge of two to $\Phi$ and treat the $\kappa H \Phi^\dagger H'$ term as an explicit B-L breaking by two units.
Since the field value of the inflaton approaches the Planck scale during inflation, the trilinear scalar term is subdominant compared to other Planck-scale suppressed effective operators and will only matter when the field value has decreased due to the cosmic expansion. Even worse, during reheating, the trilinear coupling can lead to oscillations of the scalar condensate instead of a rotation, manifesting as an oscillation in the lepton asymmetry spoiling the mechanism unless we set \cite{Barrie:2021mwi,Barrie:2022cub}
\begin{align}\label{eq:kapp}
    |\kappa| < 10^{-18}M_\text{Pl.}\simeq \mathcal{O}(\SI{10}{\giga\electronvolt}).
\end{align}
This bound is far stronger than the most naive estimate for $|\kappa|$  using the sub-eV-scale vev $v_\Phi$ in \eqref{eq:vevPhi} and $\mu_\Phi<M_\text{Pl.}$ together with $v'$ in \eqref{eq:vprime}
\begin{align}
    |\kappa| < 10^{-6}\;M_\text{Pl.}\left(\frac{v_\Phi}{m_\nu}\right)\left(\frac{10^{12}\;\text{GeV}}{v'}\right),
\end{align}
that is compatible with our previous assumption $|\kappa|\ll \mu_\Phi$.
Additionally, for the asymmetry generation, an operator of dimension larger than four is needed so that the produced lepton number is conserved during reheating \cite{Barrie:2021mwi,Barrie:2022cub}. Consequently, we consider the following dimension five operator
\begin{align}\label{eq:dim5}
    \frac{\lambda_5}{M_\text{Pl.}}\left(H \Phi^\dagger H' \pm H'^\dagger \Phi H^\dagger \right)\left( H^\dagger H \pm  H'^\dagger H'\right),
\end{align}
which conserves the discrete exchange symmetry if both signs are the same. The origin of this operator will be elucidated in appendix \ref{sec:AppC}. If we have opposite signs in both brackets, the operator violates the discrete exchange symmetry  explicitly and we might be able to use it as a bias term to remove the domain walls.  In the following we stick to the symmetry-conserving case and use plus signs following \cite{Barrie:2021mwi,Barrie:2022cub} so that the dimension five term is $\propto \cos(\phi)$. Up to mixing angles between the scalars during inflation and order one factors the lepton asymmetry at the end of inflation turns out to be \cite{Barrie:2021mwi,Barrie:2022cub}
\begin{equation}\label{eq:ADsym}
    n_{\text{L}\;\text{end}} \simeq -2 \; \lambda_5\; \rho_\text{end}^3 \;\frac{\sin\left(\phi_ 0\right)}{\sqrt{3\tilde{\lambda}}},
\end{equation}
where the factor of two takes the B-L charge of $\Phi$ into account, $\tilde{\lambda}$ is the effective quartic self-coupling of the inflaton, $\rho_\text{end}\simeq \mathcal{O}(M_\text{Pl.})$ the field value of the inflaton at the end of inflation and $\phi_0$ is the initial phase of $\rho_\Phi e^{i\phi }$. Taking into account the redshifting of the lepton asymmetry during reheating and the sphaleron redistribution coefficient, one finds that the baryon to photon ratio today can be explained for $  \lambda_5 \sin\left(\phi_0\right)/\sqrt{3\tilde{\lambda}}\simeq \mathcal{O}(10^{-16})$ \cite{Barrie:2021mwi,Barrie:2022cub}. Evidently, small values of $\lambda_5$ and $\phi_0$ are needed which also suppress isocurvature fluctuations \cite{Barrie:2021mwi,Barrie:2022cub} and a small $\lambda_5$ is necessary anyway to not spoil inflation from the non-minimal coupling. We assume a thermalized bidoublet after reheating. In order to efficiently transmit the asymmetry from the bidoublet to the leptons we have to require that the decay width $\Gamma(\Phi\rightarrow l l ')$ is larger than the decay width to scalars $\Gamma(\Phi\rightarrow H H ')$ leading to
\begin{align}\label{eq:mubound2}
    \mu_\Phi < 10^{21}\;\text{GeV}\;  Y_\nu^2 \left(\frac{v'}{10^9\;\text{GeV}}\right) \left(\frac{\SI{0.1}{\electronvolt}}{m_\nu}\right).
\end{align}
Moreover the interaction $\Phi\leftrightarrow H H '$ should be out of equilibrium  and we estimate that $ \Gamma(  H H '\rightarrow \Phi)|_{T=\mu_\Phi}\simeq \Gamma(\Phi\rightarrow H H ')|_{T=\mu_\Phi} $ is slower than the Hubble rate $H(T)$ at $T=\mu_\Phi$ as long as
\begin{align}
    v_\Phi < \SI{10}{\mega\electronvolt} \left(\frac{v'}{10^9\;\text{GeV}}\right)  \sqrt{\frac{\SI{5e10}{\giga\electronvolt}}{\mu_\Phi}}.
\end{align}
Of course we also need to ensure that the process $l l' \leftrightarrow H H'$ via off-shell $\Phi$ does not thermalize, which sets a weaker bound compared to \eqref{eq:mubound2}
\begin{equation}
    \mu_\Phi < 10^{25}\;\text{GeV}\; \left(\frac{v'}{10^9\;\text{GeV}}\right)^2 \left(\frac{\SI{0.1}{\electronvolt}}{m_\nu}\right)^2.
\end{equation}
Let us note that the cosmological history in \cite{Barrie:2021mwi,Barrie:2022cub} has an inflationary reheating temperature of $\mathcal{O}(10^{14}\;\text{GeV})$, which is in conflict with the requirement  \eqref{eq:TRH} for the absence of charged mirror leptons and quarks. That means we either need an additional mechanism to suppress reheating the mirror sector  via  asymmetric reheating  \cite{Craig:2016lyx,Chacko:2016hvu} or simply a different scenario, where $\Phi$ is not the inflaton and its large initial field value has a different origin during inflation \cite{Dine:1995uk,Co:2020dya}.
That way we can sequester the asymmetric reheating from the Affleck-Dine dynamics. Before we close, we would like to mention that there exists no obvious dark matter candidate in this model: The neutral component of $\Phi$ can decay to neutrinos or gauge bosons and will in general not be long-lived enough due its large mass. The heavy $H', Z'$ are also not long-lived enough as they couple  to all fermions (via mixing). Stable mirror quarks could form electrically neutral dark mesons  after the QCD phase transition \cite{Kang:2006yd}, however here we assume that the mirror  sector is never populated to begin with or heavily diluted (see \eqref{eq:TRH}). Therefore dark matter has to come from a separate dark sector.

\section{Conclusion}
\noindent We have presented a high scale Dirac neutrino mass model in the Type II Seesaw spirit, that has the same scalar spectrum of neutral, singly- and doubly-charged scalars as the original Majorana Type II Seesaw. This idea was implemented by introducing a bidoublet scalar in a mirror sector model with the gauge group $\text{SU}(2)_\text{L} \otimes \text{SU}(2)'\otimes \text{U}(1)_\text{X}$, where we identify the mirror neutrinos as the Dirac partners of the SM neutrinos. It was shown that the bidoublet is compatible with the discrete symmetry based solution to the strong CP problem, which was the motivation behind the mirror sector to begin with. The super-heavy bidoublet does not lead to any observable signatures for collider or other terrestrial experiments. However it might have played a role in the early universe as the source of the matter-antimatter asymmetry via either the non-thermal decay scenario or Affleck-Dine  Dirac leptogenesis. 

\acknowledgments
\noindent 
We are grateful to Andreas Trautner and Alessandro Valenti for many useful comments including, but not limited to, subtle aspects of discrete symmetries  as well as to Saurabh Nangia for valuable feedback on earlier iterations of this manuscript. This work benefited from the use of \verb|PackageX| \cite{Patel:2015tea,Patel:2016fam} and we sincerely hope that this  tool and its codebase will live on in one way or another.

\appendix
\onecolumngrid
\title{The Appendix}
\section{Scalar Sector}
\subsection{Scalar Potential}\label{sec:AppA}
\noindent 
The most general scalar potential satisfying the  discrete exchange symmetry of \cite{Hall:2018let,Dunsky:2019api}  defined in \eqref{eq:exch}  reads:
\begin{align}
  V_\Phi &= \mu_{\Phi}^2 \text{Tr}\left(\Phi^\dagger \Phi\right) 
     + \lambda_{\Phi} \text{Tr}\left(\Phi^\dagger \Phi\right)^2\label{eq:VPhi}   \\
    V_H &= - \mu_H^2\left( H^\dagger H + H'^\dagger H'\right)
    + \lambda_H \left( H^\dagger H + H'^\dagger H'\right)^2
    + \lambda'   H^\dagger H\;H'^\dagger H'\label{eq:VH}\\
    V_{H\Phi} &= \kappa \left( H \Phi^\dagger H' + H'^\dagger \Phi H^\dagger \right) 
        + \lambda_{H\Phi} \left(H \Phi^\dagger \Phi H^\dagger +H'^\dagger \Phi \Phi^\dagger H'   \right) + \alpha_{H\Phi}  \left( H^\dagger   H + H'^\dagger H'\right)  \text{Tr}\left(\Phi^\dagger \Phi\right)\label{eq:VHPhi}
\end{align}
The scalar sector involving only $H,H'$ in \eqref{eq:VH} has an approximate, global $\text{SU}(4)$ custodial symmetry so that we can embed the doublets in its fundamental representation
\begin{align}
    \mathcal{H} \equiv \begin{pmatrix} H\\ H'\end{pmatrix}.
\end{align}
Using this parameterization it is evident, that the terms $\propto \lambda', \kappa$ in \eqref{eq:VH} and \eqref{eq:VHPhi} explicitly violate  the custodial  $\text{SU}(4)$ symmetry. We take $\mu_{\Phi}^2\gg\mu_H^2>0$. All couplings are real as a consequence of the discrete exchange symmetry \eqref{eq:exch} and because $\Phi$ has an abelian charge under $\text{U}(1)_\text{X}$. If the bidoublet was uncharged, the potential would depend on both  $\Phi$ and $\tilde{\Phi}\equiv-\sigma_2 \Phi^* \sigma_2$ leading to explicit CP-violation via terms like 
\begin{align}
    \alpha_{H\Phi2}\left(H^\dagger H\; \text{Tr}\left(\tilde{\Phi}^\dagger \Phi\right)+H'^\dagger H'\; \text{Tr}\left(\tilde{\Phi} \Phi^\dagger\right)   \right) + \text{h.c.}
\end{align}
with a complex coupling $\alpha_{H\Phi2}$. For the charged bidoublet case loop diagrams do not regenerate the explicit CP-violating scalar couplings unlike the case of the minimal LRSM \cite{Kuchimanchi:2014ota}.

\subsection{Minimization of the scalar potential: Original Higgs-Parity model}\label{sec:minima}
\begin{figure}[t]
    \centering
    \includegraphics[width=0.45\textwidth]{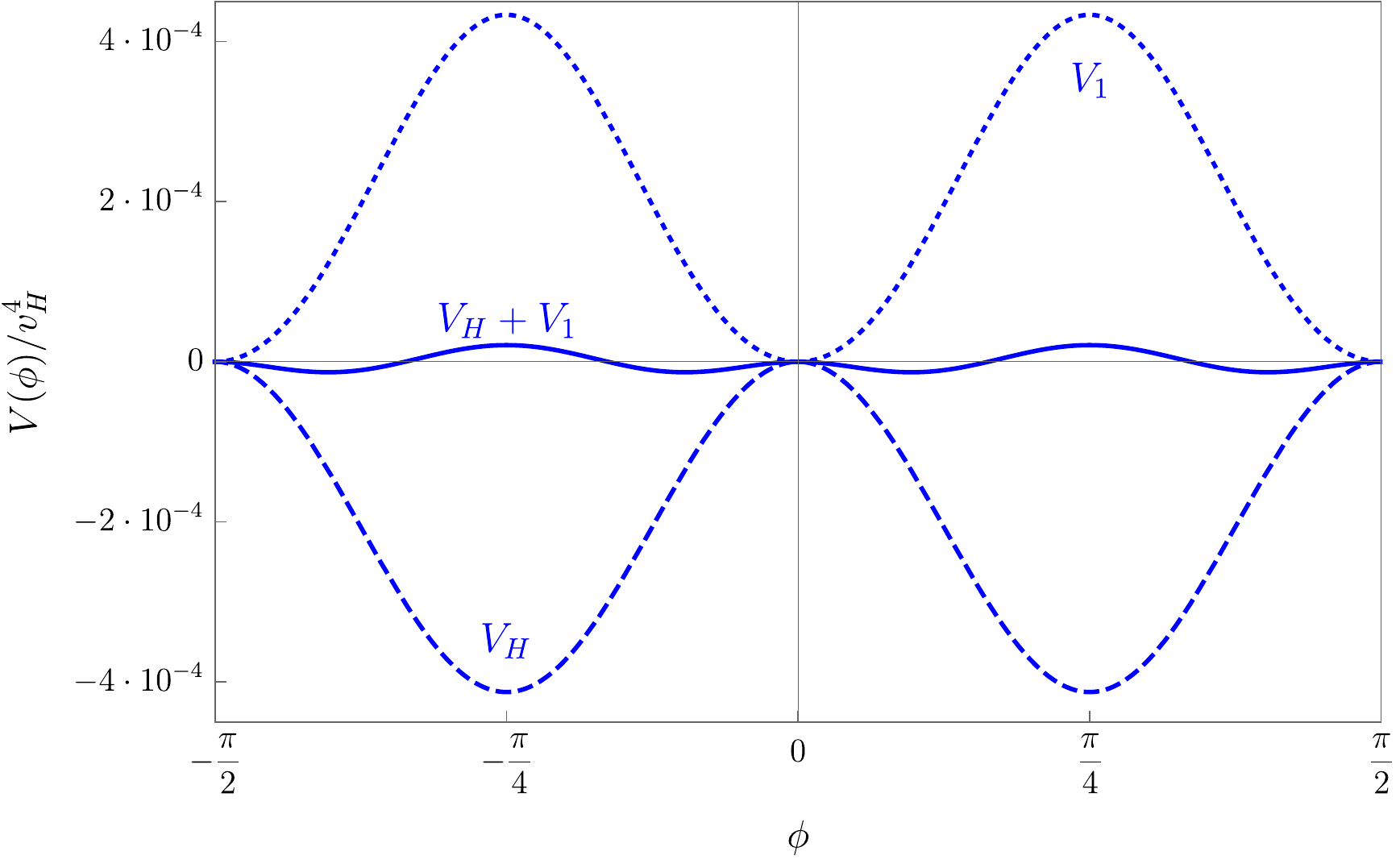}
    \includegraphics[width=0.45\textwidth]{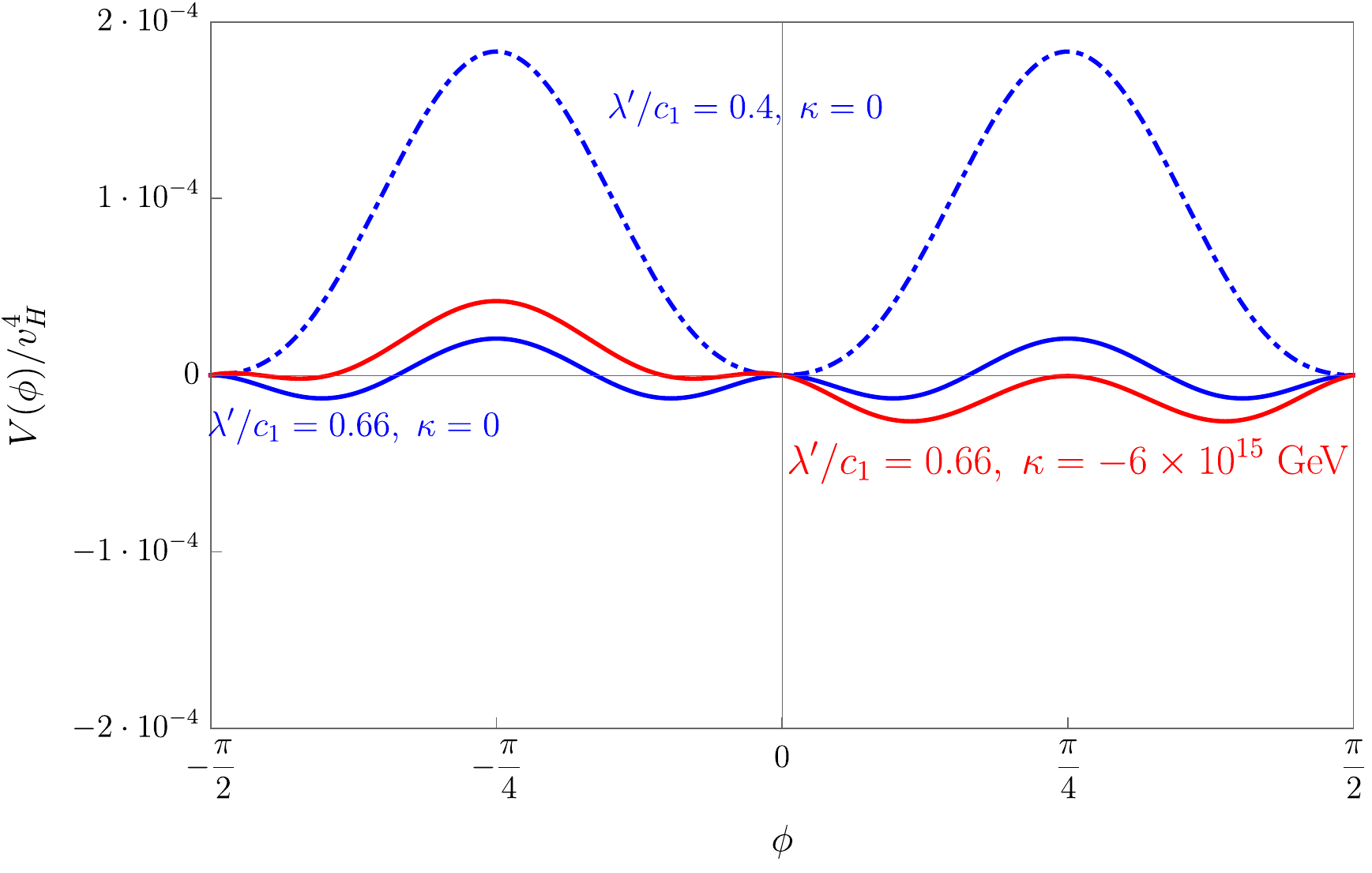}
    \caption{Plot of the contributions to the scalar potential without a bidoublet \textit{(left)} and with a bidoublet \textit{(right)}. For the sake of visibility and illustration we chose $v_H = 10^{10}\;\text{GeV}$, $v_\Phi = 1\;\text{GeV}$, $c_1=-10^{-2}$ and $\kappa = -6\times10^{15}\;\text{GeV}$, which do not correspond to phenomenologically viable parameters. On the left hand side we fixed $\lambda'/c_1=0.66$ and we varied this combination of parameters on the right hand side. Note that we scaled the vertical axis  differently  in both plots.
    Realistic parameters would lead to minima at $|\phi| \simeq 10^{-7}-10^{-10}$ and in practise the phenomenologically required small value  of $|\kappa|$, e.g. $|\kappa|<\SI{10}{\giga\electronvolt}$ from \eqref{eq:kapp} in the main text, only has a negligible impact on the value of $\phi$.  }
    \label{fig:onlyplot}
\end{figure}
\noindent 
Here we explore the minima of the scalar potential in the electrically neutral directions. For the other directions see the next section \ref{sec:AppB}. We begin our discussion of the minimization of the scalar potential in \eqref{eq:VPhi}-\eqref{eq:VHPhi} by introducing the notation
\begin{align}\label{eq:param}
    v_H\equiv \sqrt{v^2+v'^2},\quad \sin(\phi) \equiv \frac{v}{v_H},\quad \cos(\phi)\equiv \frac{v'}{v_H}.
\end{align}
The phenomenologically required vacuum structure is $v'\neq 0 \gg v \neq 0$. This together with $v\neq v'$ implies that $\phi \in (0,\frac{\pi}{4})$. For the observed value of $v=246\;\text{GeV}$ and the required $v'  \simeq \left(10^9-10^{12}\right)\;\text{GeV}$ (see \eqref{eq:vprime}) we have $0<\phi \simeq 10^{-7}-10^{-10} \ll1$ and $v_H\simeq v'$. 

\noindent First we will summarize the results of \cite{Hall:2018let,Dunsky:2019api} for the scalar potential involving only $H$ and $H'$, before discussing the impact of the bidoublet. The vacuum potential reads
\begin{align}\label{eq:pot2}
    V_H = \frac{v_H^2}{32}\left(-16 \mu_H^2 + v_H^2 \left(8 \lambda_H + \lambda' (1-\cos(4\phi)) \right)  \right)
\end{align}
and the minimization conditions are found to be
\begin{align}
    \frac{\partial V_H}{\partial v_H} &=  \frac{v_H}{8}\left(-8 \mu_H^2 + v_H^2 \left(8 \lambda_H + \lambda' (1-\cos(4\phi)) \right)  \right),\quad \text{and} \quad 
     \frac{\partial V_H}{\partial \phi} = \frac{v_H^4}{8}\lambda' \sin(4\phi).
\end{align}
 The second condition has the solutions $\phi = (0,\; \pi/2,\; \pi/4)$ corresponding to $(v=0,\; v'=0,\; v=v')$, where in the first (second) case we have $v'\neq 0\;(v\neq0)$. This essentially happens, because the potential for $\phi$ has periodicity of $\pi/2$ and a reflection symmetry \cite{Jung:2019fsp} owing to the Higgs-parity defined in \eqref{eq:exch}. 
 While the custodial-symmetry-breaking and Higgs Parity conserving interaction $ \lambda'  H^\dagger H\;H'^\dagger H'$ allows us to find an asymmetric vacuum with $v=0\;\text{and}\; v'\neq0$, it does not suffice in order to also break the electroweak gauge symmetry. To realize $v\neq 0$ we need a separate source of custodial symmetry violation, that slightly tilts the potential even further. Yukawa and gauge interactions break the custodial symmetry explicitly and these effects are communicated to the scalar potential via quantum corrections encoded in the one-loop Coleman-Weinberg potential \cite{PhysRevD.7.1888}
 \begin{align}
     V_{1} &= c_1 \left( \left( H^\dagger   H \right)^2 \log\left(\frac{\left|H\right|}{\mu}\right) +
     \left( H'^\dagger   H' \right)^2 \log\left(\frac{\left|H'\right|}{\mu}\right)\right),\quad
     c_1 \equiv -\frac{3}{8\pi^2}Y_t^4 + \frac{3}{128\pi^2}\left(g^2+g'^2\right)^2+ \frac{3}{64\pi}g^4.
 \end{align}
 The negative contribution from the top quark Yukawa is the dominant one, which is why $c_1<0$.
 If we plug in the values of the Yukawa and gauge couplings around the weak scale as an estimate we find $|c_1|<10^{-2}$. In terms of the parameterization \eqref{eq:param} this potential reads for a renormalization scale of $\mu =v_H$ \cite{Jung:2019fsp}
 \begin{align}
     V_1 &= c_1 v_H^4\left(\cos(\phi)^4\log\left(\cos(\phi)\right)+\sin(\phi)^4\log\left(\sin(\phi)\right)\right)\\
     &= \frac{c_1 v_H^4}{4}\left(\frac{25-24\log(2)}{96}\cos(4\phi)-\frac{1}{240}\cos(8\phi)
     -\frac{1}{2240}\cos(12\varphi) -\frac{1}{10080}\cos(16\varphi)    +\mathcal{O}\left(\cos(20\phi)\right)\right)\label{eq:pot3}.
 \end{align}
 Following \cite{Jung:2019fsp} we only include the terms up to $8\phi$ as we find the rest to be negligible due to numerically small coefficients.  A partial cancellation between the $\cos(4\phi)$ terms in \eqref{eq:pot2} and \eqref{eq:pot3} will allow us to find a viable  solution $0<\phi\ll\frac{\pi}{4}$. This is also why we only display the leading order coefficients of a Fourier expansion in $\cos(n \;4\phi)$ with $n\in\mathbb{N}$.  The new minimization conditions are found to be 
 \begin{align}
    \frac{\partial V}{\partial v_H} &=  \frac{1}{480}\left(60v_H\left(-8 \mu_H^2 + v_H^2 \left(8 \lambda_H + \lambda'  \right)\right)- v_H^3\left(5\cos(4\phi)\left(12\lambda' + c_1 (24 \log(2)-25)\right)-2 c_1 \cos(8\phi) \right) \right),\\
     \frac{\partial V}{\partial \phi} &=  \frac{v_H^4}{480}\left( 60 \lambda'+8 c_1 \cos(4\phi) + 5 c_1 (24 \log(2)-25)\right)\sin(4\phi)\label{eq:brack}.
\end{align}
When solving for $\phi$ one has two solutions: Either $\sin(4\phi)=0$, which implies the solutions $\phi = (0,\;\pi/2,\;\pi/4)$ for the unwanted set of either  partially unbroken or symmetric vacua. Else, the second factor in \eqref{eq:brack} has to be zero itself for a solution with non-zero $\phi\ll1$. We can solve this equation to find the required $\lambda'$ for a minimum, that can accommodate the  input parameter $\phi$
\begin{align}\label{eq:lambdaaa}
    \lambda' = \frac{c_1}{60} (125 - 60\log(4) -8\cos(4\phi))\overset{\phi\ll1}{\simeq} 0.56 \;c_1,
\end{align}
and substitute this into the first minimization condition to obtain
\begin{equation}\label{eq:minvH}
    v_H = \frac{\mu_H}{\sqrt{\lambda_H + \frac{c_1}{16}\left(\frac{645}{10}-4\log(2)-\frac{4}{15}\cos(4\phi)+\frac{1}{15}\cos(8\phi)\right)  }}\overset{\phi\ll1}{\simeq} \frac{4\sqrt{10}\mu_H}{\sqrt{160\lambda_H + c_1 ( 41-40\log(2))}}\overset{|c_1|\ll1}{\simeq}\frac{\mu_H}{\sqrt{\lambda_H}}.
\end{equation}
The Higgs mass and self coupling at the scale $\mu=v_H$ are found to be \cite{Jung:2019fsp}
\begin{align}
    m_h^2\simeq  -\left(\lambda'-\frac{c_1}{2}\right)v_H^2, \quad \text{and} \quad \lambda_h(\mu=v_H)=\frac{c_1}{16}\lesssim0.
\end{align}
The lightness of $m_h$ with respect to the high scale $v_H\simeq v'$ is related to the tuning $\lambda'\simeq c_1 / 2 \lesssim 0$ \cite{Dunsky:2019api} in \eqref{eq:lambdaaa}, which manifests the previously mentioned partial cancellation. If $\lambda' / c_1$ stays between 0.5 and 0.81, the unwanted values with $\phi = (0,\;\pi/4)$ are actually maxima of the scalar potential, as can be seen from its second derivative \cite{Jung:2019fsp}. The sign of $\phi$ is in general undefined and the solution to \eqref{eq:brack} reads
\begin{align}
  \phi = \pm \frac{1}{4}\text{arccos}\left(\frac{5}{8}\left(25-24\log(2) -12 \frac{\lambda'}{c_1}\right)\right),
\end{align}
which is a consequence of the reflection symmetry of the potential. Since a physically sound vev must satisfy $v>0$, we have to impose $\phi>0$. We illustrate the previously discussed partial cancellation between the tree level potential $V_H$ and the Coleman-Weinberg terms $V_1$ on the left side of figure \ref{fig:onlyplot}. One can see that there are two symmetric non-zero minima with $|\phi| < \pi/4$. On the right hand side of the aforementioned figure we plotted the potential for different choices of $\lambda'/c_1$ and one can clearly observe that the non-zero values of $|\phi| < \pi/4$ require $\lambda'/c_1\gtrsim 1/2$. For the plots we used unrealistic parameters for the sake of being able to see the minima of $\phi$ between 0 and $\pm\pi/4$.  Realistic parameters would lead to minima at $|\phi| \simeq 10^{-7}-10^{-10}$.

\subsection{Minimization of the scalar potential: Inclusion of the bidoublet}\label{sec:minima2}
\noindent Next we introduce the couplings to the bidoublet
\begin{align}\label{eq:vphiphi}
   V_\Phi &= \frac{v_\Phi^2}{2}\left(\mu_\Phi^2+\frac{\lambda_\Phi}{2}v_\Phi^2\right), \quad
   V_{H\Phi} = v_\Phi v_H^2 \left(\frac{v_\Phi}{4}\left(\alpha_{H\Phi}+\lambda_{H\Phi}\right) +\frac{\kappa}{2\sqrt{2}}\sin(2\phi) \right),
\end{align}
 where we see that only the trilinear term $\propto \kappa$ depends on $\phi$ and thus violates the custodial symmetry. Furthermore, since this term is $\propto \sin(2\phi)$ the scalar potential for $\phi$ no longer has the periodicity $\pi/2$. The modified minimization conditions read
 \begin{align}
    \frac{\partial V}{\partial v_H} &=  \frac{1}{480}\left(60v_H\left(-8 \mu_H^2 + v_H^2 \left(8 \lambda_H + \lambda'  \right)\right)- v_H^3\left(5\cos(4\phi)\left(12\lambda' + c_1 (24 \log(2)-25)\right)-2 c_1 \cos(8\phi) \right) \right)\\
    &+ \frac{v_\Phi v_H}{2} \left( v_\Phi \left(\alpha_{H\Phi}+\lambda_{H\Phi}\right) +\sqrt{2} \kappa\sin(2\phi) \right)\nonumber,\\
    \frac{\partial V}{\partial v_\Phi} &= v_\Phi\left(\mu_\Phi^2+\lambda_\Phi v_\Phi^2 + \left(\alpha_{H\Phi}+\lambda_{H\Phi}\right) v_H^2 \right) + \frac{v_H^2}{2\sqrt{2}}  \kappa\sin(2\phi)  \label{eq:TRIN} ,\\
     \frac{\partial V}{\partial \phi} &=  \frac{v_H^4}{480}\left( 60 \lambda'+8 c_1 \cos(4\phi) + 5 c_1 (24 \log(2)-25) + 120\sqrt{2}\frac{ v_\Phi \kappa}{v_H^2} \frac{1}{\sin(2\phi)}\right)\sin(4\phi).
\end{align}
The required $\lambda'$ for values of $\phi \neq (0,\;\pi/2,\;\pi/4)$ that minimize the potential in the  $\phi$-direction is found to be
\begin{align}\label{eq:solprime}
    \lambda' = \frac{c_1}{60} (125 - 60\log(4) -8\cos(4\phi)) -2\sqrt{2} \frac{v_\Phi \kappa}{v_H^2}\frac{1}{\sin(2\phi)}  \overset{\phi\ll1}{\simeq} 0.56 \;c_1 -\frac{\sqrt{2}}{\phi} \frac{v_\Phi \kappa}{v_H^2}.
\end{align}
Note that the trilinear term $ v_\Phi  \kappa  v_H^2 \sin(2\phi)$ in  \eqref{eq:vphiphi} breaks the reflection symmetry and biases the vaccuum in the direction $\phi>0\;(\phi<0)$ for $\kappa<0\;(\kappa>0)$ (analogous to the sign of $v_\Phi$ for a Type II Seesaw). However in practise this contribution is suppressed as $v_\Phi \kappa /v_H^2$ compared to  $V_H+V_1$, so that we would need to take large (and phenomenologically excluded) values of $\kappa$   to select a sign for $\phi$. This was illustrated on the right side of figure \ref{fig:onlyplot} and one sees the deeper minimum $\phi>0$ for the unrealistically large $\kappa=-\SI{6e15}{\giga\electronvolt}\neq 0$ in red. The next paragraphs explain, why we can not make $|\kappa|$ arbitrarily large and phenomenologically we need a small value of $|\kappa| < 10\;\text{GeV}$ (see \eqref{eq:kapp} in the main text) anyway. 
We find that $v_H$ is determined to be
\begin{equation}\label{eq:solvH}
    v_H = 4\frac{\sqrt{30\mu_H^2 -15 v_\Phi\left( v_\Phi \left(\alpha_{H\Phi}+\lambda_{H\Phi}\right) +\sqrt{2} \kappa\sin(2\phi) \right)  }}{\sqrt{60(8\lambda_H+\lambda')-2c_1 \cos(8\phi) - 5 \cos(4\phi)\left(12\lambda'+c_1 (24 \log(2)-25)\right)}}.
\end{equation}
If the first term $30\mu_H^2$ dominates over the contribution $\propto v_\Phi^2,  v_\Phi \kappa$,
the previously determined minimum in \eqref{eq:minvH} is still valid. Once the  contribution $\propto v_\Phi^2, v_\Phi \kappa$ takes over,  a deeper minimum starts to appear and the vev $v_H$ is actually induced by $v_\Phi$ (instead of the other way around for a Type II Seesaw).   To study the implications of $v_\Phi$ on $\phi, v_H$  requires  finding, which value of $v_\Phi$ solves \eqref{eq:TRIN}.  If we were to switch off all bidoublet couplings to the other scalars and set $\mu_\Phi^2<0$, we expect $v_\Phi = |\mu_\Phi| / \sqrt{\lambda_\Phi}$ as usual. Generally speaking this relation will be modified by the vevs of the other Higgses as well, because the the trilinear coupling $\kappa$ and the full solution to \eqref{eq:TRIN} can only be found numerically. In the following we either fix $v_\Phi$ via the Type II Seesaw scheme used in the main text, or use it as a free parameter in order to find the conditions for unwanted symmetric  or deeper minima.

\subsubsection{Induced bidoublet vev a la Type II Seesaw}\label{sec:a}
\noindent 
Here we take the vev $v_\Phi$ to be induced by $v_H$. This means, we assume $\mu_\Phi^2>0$ and furthermore,   that $\mu_\Phi^2 \gg \lambda_\Phi v_\Phi^2 + \left(\alpha_{H\Phi}+\lambda_{H\Phi}\right) v_H^2 $, so that \eqref{eq:TRIN} is approximately solved by
\begin{align}\label{eq:vevagain}
    v_\Phi \simeq  -\frac{\kappa\;v_H^2}{2\sqrt{2}\mu_\Phi^2} \sin(2\phi) =  -\frac{\kappa\;v\;v'}{\sqrt{2}\mu_\Phi^2}.
\end{align}
In the limit $|\kappa|\ll v_H\ll \mu_\Phi$  this vev will essentially be the smallest scale in the potential. The value of $\lambda'$ required for a given $\phi$ in \eqref{eq:solprime} then reads
\begin{align}
    \lambda'_\text{eff.}\equiv \lambda' - \frac{\kappa^2}{\mu_\Phi^2} = \frac{c_1}{60} (125 - 60\log(4) -8\cos(4\phi))  \overset{\phi\ll1}{\simeq} 0.56 \;c_1.
\end{align}
We see, that integrating out the super-heavy bidoublet just shifts the coupling $\lambda'$ via a threshold correction, as was mentioned above of \eqref{eq:thr} in the main text. Since we expect $|\kappa| \ll \mu_\Phi$ by many orders of magnitude (see \eqref{eq:estimate} and the discussion below), it is safe to take $\lambda'_\text{eff.}\simeq \lambda'$ and the previously determined minimum for $\phi$ in \eqref{eq:lambdaaa} is still valid. 
In order to avoid deeper minima  than $v_H$  in \eqref{eq:minvH} we have to require that the numerator in \eqref{eq:solvH} satisfies
\begin{align}
   2 \mu_H^2 \gg v_\Phi\left( v_\Phi \left(\alpha_{H\Phi}+\lambda_{H\Phi}\right) +\sqrt{2} \kappa\sin(2\phi) \right) \simeq \frac{\kappa^2 v_H^2 }{2\mu_\Phi^2}\sin(2\phi)^2,
\end{align}
where we used $\mu_\Phi^2 \gg  \left(\alpha_{H\Phi}+\lambda_{H\Phi}\right) v_H^2 $ in the last step. Expanding for small $\phi$ and setting $v_H\simeq \mu_H /\sqrt{\lambda_H}$ turns this into 
\begin{align}
    \lambda_H \gg \phi^2 \frac{\kappa^2}{\mu_\Phi^2}.
\end{align}
Since we assume $\lambda_H = \mathcal{O}(1)$ and again stress that $\phi, \;|\kappa|/ \mu_\Phi \ll 1$, we do not need to worry about deeper minima for $v_H$ with the super-light $v_\Phi$ we consider in \eqref{eq:vevagain}.

\subsubsection{General bidoublet vev}
\noindent As discussed earlier, it is in general not possible, to obtain a full analytic expression for $v_\Phi$. This is why, we take it as a free parameter and in the following   make no assumption about its relative size compared to $v_H, \mu_\Phi$ and $\kappa$. Inspecting \eqref{eq:solprime} reveals, that  the $1/\sin(2\phi)$ factor can become large for the required small $\phi\ll1$ and if it is not cancelled by $v_\Phi \kappa /v_H^2$, it might happen, that this term becomes larger than the perturbative limit for $\lambda'$  of $4\pi$. We therefore require that
\begin{equation}
    \left|0.56 \;c_1 -\frac{\sqrt{2}}{\phi} \frac{v_\Phi \kappa}{v_H^2}\right| < 4 \pi,
\end{equation}
where the absolute value takes into account the in general undetermined sign of $\kappa$ and the fact that $c_1<0$.  Assuming the bidoublet contribution is larger than the Coleman-Weinberg piece $\propto c_1$, one finds that this condition implies
\begin{align}\label{eq:bound1}
    v_\Phi |\kappa| <    2\sqrt{2} \pi\; v\; v'.
\end{align}
In other words, if we make $v_\Phi |\kappa|$ larger than the input parameters $v\;v'$, then the corrections from the bidoublet vev will spoil the partial cancellation between the $\lambda'$ and the  Coleman-Weinberg terms $\propto c_1$ responsible for the correct asymmetric vacuum $v\neq 0\ll v'$. The $\sin(2\phi)$ term coming from the coupling to the bidoublet in \eqref{eq:vphiphi} is responsible for this effect, since the aforementioned partial cancellation involves the $\cos(4\phi)$ terms. This is in agreement with the findings of \cite{Hall:2018let}, who arrived at the conclusion, that the vev of an additional bidoublet can not contribute significantly to electroweak symmetry breaking. Consequently we are forced to have a small $ v_\Phi |\kappa|$. Avoiding a deeper minimum from \eqref{eq:solvH} than the $v_H$ in \eqref{eq:minvH} requires 
\begin{align}
     \mu_H^2 \gg \frac{1}{\sqrt{2}} v_\Phi \kappa\sin(2\phi),
\end{align}
where we assumed that $\alpha_{H\Phi}+\lambda_{H\Phi}$ is negligible so that we can focus on $\kappa$. This bound can be re-expressed as 
\begin{align}
   v_\Phi |\kappa| \ll \frac{\lambda_H}{\sqrt{2}}\frac{v'^3}{v}.
\end{align}
We find that this constraint is weaker than \eqref{eq:bound1} for $v\ll v'$, meaning that taking $ v_\Phi |\kappa|$ to be large will first destroy the misalignment of vacua (leading to $v=v'$ or $v=0$) before it leads to deeper minima in $v'$. The smallness of the induced $v_\Phi$ in the Type II Seesaw of \eqref{eq:vevagain} automatically avoids these problems in the limit  $|\kappa|\ll \mu_\Phi$.

\subsection{Sufficient conditions for vacuum stability}\label{sec:AppB}
\noindent 
For vacuum stability at large field values only the quartic terms are important.
Following reference \cite{Arhrib:2011uy} we define 
\begin{align}
    r^2 &\equiv  H^\dagger H + H'^\dagger H' +\text{Tr}\left(\Phi^\dagger \Phi\right),\quad
    r^2 \cos(\gamma) \equiv  H^\dagger H + H'^\dagger H',\quad
    r^2 \sin(\gamma) \equiv \text{Tr}\left(\Phi^\dagger \Phi\right),\\
    x &\equiv \frac{ H^\dagger H\;H'^\dagger H'}{\left(H^\dagger H + H'^\dagger H'\right)^2},\quad
    y \equiv \frac{H \Phi^\dagger \Phi H^\dagger +H' \Phi \Phi^\dagger H' }{\left(H^\dagger H + H'^\dagger H'\right)\;\text{Tr}\left(\Phi^\dagger \Phi\right)}.
\end{align}
One can show that 
\begin{align}
    0\leq x \leq \frac{1}{2}, \quad 0\leq y\leq 1.
\end{align}
Using this parameterization we employ the co-positivity criteria of \cite{Kannike:2012pe} to find
\begin{align}
    \lambda_\Phi>0,\quad
    \lambda_H + x \lambda' >0,\quad
    \alpha_{H\Phi}+  y \lambda_{H\Phi} + 2 \sqrt{\lambda_\Phi \left(\lambda_H + x \lambda'\right)}>0.
\end{align}
Note that $x,y$ may not be independent parameters \cite{Bonilla:2015eha}, however we will ignore this complication for our first estimate. This is the reason why we only find the sufficient but not the necessary criteria for vacuum stability. A more refined analysis along the lines of \cite{Bonilla:2015eha,Moultaka:2020dmb,Han:2022ssz} is required to treat the  general case. Our preliminary investigation did not find deeper electric charge-breaking minima compared to the charge-conserving ones in \eqref{eq:ansatz}, which can occur for models with trilinear scalar couplings \cite{Barroso:2005hc}, and we do not expect them due to the smallness of the trilinear coupling $\kappa$ (see the previous paragraph and the discussion below \eqref{eq:estimate}). A full numerical analysis is beyond the scope of this work.

\section{Origin of the dimension five operator for Affleck-Dine leptogenesis}\label{sec:AppC}
\noindent The discrete exchange symmetry  conserving effective operator in \eqref{eq:dim5} (same signs in each bracket) can be realized by including a real singlet either even ($\sigma(t,\vec{x}) \rightarrow \sigma(t,-\vec{x})$) or odd ($\sigma(t,\vec{x}) \rightarrow -\sigma(t,-\vec{x})$) under the symmetry \eqref{eq:exch}. This adds the following terms to the scalar potential
\begin{align}
    V_\sigma &= \mu_\sigma^2 \sigma^2 + \lambda_{3\sigma} \sigma^3 + \lambda_{4\sigma} \sigma^4,\\
    V_\text{int.} &= \kappa_\sigma \sigma \left(H^\dagger H \pm  H'^\dagger H'\right) + \lambda_{\sigma H \Phi H'}\;\sigma \left( H \Phi^\dagger H' \pm  H'^\dagger \Phi H^\dagger\right)\\
    &+ \lambda_{H\sigma} \;\sigma^2 \left( H^\dagger H + H'^\dagger H'\right) + \lambda_{ \Phi\sigma}\;  \sigma^2 \text{Tr}\left(\Phi^\dagger \Phi\right)  ,
\end{align}
where the $+(-)$ sign applies to the even (odd) case and furthermore the term $\lambda_{3\sigma} \sigma^3$ is absent for odd $\sigma$. Note that all of the above couplings are real. If we assume the $\sigma$ mass is the largest in the potential, we find after integrating it out, that the effective coupling in \eqref{eq:dim5} is given by
\begin{align}
     \frac{\lambda_5}{M_\text{Pl.}} \simeq \frac{\lambda_{\sigma H \Phi H'}\; \kappa_\sigma}{|\mu_\sigma|^2}.
\end{align}
In order for this operator to be present during the inflationary stage, where the inflaton can approach Planck scale field values, we have to impose $|\mu_\sigma|\simeq \mathcal{O}(M_\text{Pl.})$. A vev for $\sigma$ is not required to generate the right operator. For completeness let us consider the implications of a vev for the real $\sigma$:
\begin{itemize}
    \item \textbf{even case:} Since the scalar $\sigma$ is even, the discrete exchange symmetry   remains unbroken and $v_\sigma$ simply shifts $\mu_\Phi^2\rightarrow \mu_\Phi^2 + \lambda_{\Phi\sigma} v_\sigma^2,\;-\mu_H^2\rightarrow -\mu_H^2 + \kappa_\sigma v_\sigma+\lambda_{H\sigma} v_\sigma^2$ as well as $\kappa\rightarrow \kappa +\lambda_{\sigma H \Phi H'} v_\sigma$. Since we expect $v_\sigma$ to be very large we require small  couplings in order to not shift the scales too much.
    \item \textbf{odd case:} In this scenario $\sigma$ is a  pseudo-scalar that spontaneously breaks the discrete exchange symmetry and leads to different mass terms $-\mu_H^2 \pm \kappa_\sigma v_\sigma$
    for $H,H'$ effectively realizing the softly-broken parity scenario of \cite{PhysRevD.41.1286} for $v'\gg v$ mentioned in section \ref{sec:vacuum}. On top of that   when it comes to CP  the minimum of the scalar potential will be different from \eqref{eq:phase-min} and we find spontaneous CP violation with an angle for $v_\Phi$ of
    \begin{align}
      \beta_\Phi -\beta-\beta' = \text{arctan}\left(\frac{\lambda_{\sigma H \Phi H'}\;v_\sigma}{\sqrt{2}\;\kappa}\right).
    \end{align}
    This phase is negligible for the solution to the strong CP problem in section \ref{sec:strong}, because the field $\Phi$ has no direct couplings to quarks and leptonic insertions occur only at very large and thus heavily suppressed loop orders. In the symmetry-odd case there could arise a dimension five operator similar to \eqref{eq:dim6} \cite{DAgnolo:2015uqq}
    \begin{align}
              \frac{c_5^g}{\Lambda_\text{UV}} \sigma\; G_{\mu\nu} \tilde{G}^{\mu\nu},
    \end{align}
    or a correction to the quark Yukawas in \eqref{eq:quarks} of the form \cite{DAgnolo:2015uqq}
    \begin{align}
        \frac{c_5^u}{\Lambda_\text{UV}} i\sigma\left( Y_u q H^\dagger \overline{u} + Y_u' q' H'^\dagger \overline{u}'\right) +    \frac{c_5^d}{\Lambda_\text{UV}} i\sigma\left(
        Y_d q H \overline{d} +  + Y_d' q' H' \overline{d}'\right) + \text{h.c.},
    \end{align}
    which for a Planck-scale cut-off  $\Lambda_\text{UV}$ need to satisfy $c_5^{g,u,d}\; v_\sigma < 10^9\;\text{GeV}$ \cite{DAgnolo:2015uqq}  in order to comply with the experimental bound of $\overline{\theta}<10^{-10}$ \cite{PhysRevD.19.2227,CREWTHER1979123,Baker:2006ts,2016PhRvL.116p1601G}. 
\end{itemize}

\noindent The effective operator in \eqref{eq:dim5} with different signs in each bracket, violating the discrete exchange symmetry, could arise from non-perturbative quantum gravitational effects, which are expected \cite{COLEMAN1988643,GIDDINGS1988854,GILBERT1989159} to explicitly break  all global symmetries  that are not residual symmetries of gauge symmetries.

\twocolumngrid
\bibliographystyle{utphys}
\bibliography{references-1}

\end{document}